\documentclass[twocolumn]{aastex6}

\usepackage{times,color,natbib,url,amsmath}
\usepackage{graphicx,float,psfrag}
\usepackage{tabularx}
\usepackage{latexsym,amsmath,amssymb}
\usepackage{natbib}
\usepackage{verbatim}
\usepackage{multirow}
\usepackage{courier}
\usepackage{rotating}
\usepackage{longtable}
\usepackage{pbox}
\usepackage{hyperref}
\usepackage{enumitem}
\usepackage{lipsum}

\bibpunct{(}{)}{;}{a}{}{,}

\newcommand {\xmm} {\textsl{XMM-Newton}}
\newcommand {\chandra} {\textsl{Chandra}}
\newcommand {\swift} {\textsl{Swift}}

\newcommand {\nustar} {\textsl{NuSTAR}}

\def \rsun {\ifmmode$R$_{\odot}\else R$_{\odot}$}

\def \hcm {\hbox {\ifmmode $ atoms cm$^{-2}\else atoms cm$^{-2}$\fi}}

\def\approxgt{\mathrel{\hbox{\rlap{\lower.55ex \hbox {$\sim$}}
        \kern-.3em \raise.4ex \hbox{$>$}}}}
\def\approxlt{\mathrel{\hbox{\rlap{\lower.55ex \hbox {$\sim$}}
        \kern-.3em \raise.4ex \hbox{$<$}}}}

\def \arcsec {\hbox{$^{\prime\prime}$}}

\newcommand\blfootnote[1]{%
  \begingroup
  \renewcommand\thefootnote{}\footnote{#1}%
  \addtocounter{footnote}{-1}%
  \endgroup
}

\def \src{SGR~J1935$+$2154}

\begin{document}

\title{X-ray and radio observations of the magnetar \src\
  during its 2014, 2015, and 2016 outbursts}

%
%

\author{George~Younes\altaffilmark{1,2}}
\author{Chryssa~Kouveliotou\altaffilmark{1,2}}
\author{Amruta~Jaodand\altaffilmark{3,4}}
\author{Matthew~G.~Baring\altaffilmark{5}}
\author{Alexander~J.~van~der~Horst\altaffilmark{1,2}}
\author{Alice~K.~Harding\altaffilmark{6}}
\author{Jason~W.~T.~Hessels\altaffilmark{3,4}}
\author{Neil Gehrels\altaffilmark{6}$^\dagger$}\blfootnote{$^\dagger$Deceased 2017 February 6}
\author{Ramandeep~Gill\altaffilmark{7}}
\author{Daniela~Huppenkothen\altaffilmark{8,9}}
\author{Jonathan~Granot\altaffilmark{7}}
\author{Ersin~G\"o\u{g}\"u\c{s}\altaffilmark{10}}
\author{Lin~Lin\altaffilmark{11}}


\altaffiltext{1}{Department of Physics, The George Washington University, Washington, DC 20052, USA, gyounes@gwu.edu}
\altaffiltext{2}{Astronomy, Physics and Statistics Institute of Sciences (APSIS)}
\altaffiltext{3}{ASTRON, the Netherlands Institute for Radio Astronomy, Postbus 2, 7990 AA Dwingeloo, the Netherlands}
\altaffiltext{4}{Astronomical Institute Anton Pannekoek, University of Amsterdam, 1098XH, Amsterdam, the Netherlands}
\altaffiltext{5}{Department of Physics and Astronomy, Rice University, MS-108, P.O. Box 1892, Houston, TX 77251, USA}
\altaffiltext{6}{Astrophysics Science Division, NASA Goddard Space Flight Center, Greenbelt, MD 20771}
\altaffiltext{7}{Department of Natural Sciences, The Open University of Israel, 1 University Road, P.O. Box 808, Ra\'anana 43537, Israel}
\altaffiltext{8}{Center for Data Science, New York University, 726 Broadway, 7th Floor, New York, NY 10003}
\altaffiltext{9}{Center for Cosmology and Particle Physics, Department of Physics, New York University, 4 Washington Place, New York, NY 10003, USA}
\altaffiltext{10}{Sabanc\i~University, Orhanl\i-Tuzla, \.Istanbul 34956, Turkey}
\altaffiltext{11}{Department of Astronomy, Beijing Normal University, Beijing China 100875}


\begin{abstract}

We analyzed broad-band X-ray and radio data of the magnetar \src\
taken in the aftermath of its 2014, 2015, and 2016
outbursts. The source soft X-ray spectrum $<10$~keV is well described
with a BB+PL or 2BB model during all three outbursts. \nustar\
observations revealed a hard X-ray tail, $\Gamma=0.9$, extending up to
$79~$keV, with flux larger than the one detected $<10$~keV. Imaging
analysis of \chandra\ data did not reveal small-scale extended
emission around the source. Following the outbursts, the total
$0.5-10$~keV flux from \src\ increased in concordance to its bursting
activity, with the flux at activation onset increasing by a factor of
$\sim7$ following its strongest June 2016 outburst. A
\swift/XRT observation taken $1.5$ days prior to the onset of this
outburst showed a flux level consistent with quiescence. We show
that the flux increase is due to the PL or hot BB component, which
increased by a factor of $25$ compared to quiescence, while the cold
BB component $kT=0.47$~keV remained more or less constant. The 2014
and 2015 outbursts decayed quasi-exponentially with time-scales of
$\sim40$~days, while the stronger May and June 2016 outbursts showed a
quick short-term decay with time-scales of $\sim4$~days. Our Arecibo
radio observations set the deepest limits on the radio emission from a
magnetar, with a maximum flux density limit of $14$\,$\mu$Jy for the
4.6~GHz observations and $7$\,$\mu$Jy for the 1.4~GHz observations. We
discuss these results in the framework of the current magnetar
theoretical models.

\end{abstract}

\section{Introduction}
\label{Intro}

A sub-set of isolated neutron stars (NSs), dubbed magnetars, show
peculiar rotational properties with low spin periods $P$ in the range
of $2-12$ seconds and large spin down rates $\dot{P}$ of the order of
$10^{-11}-10^{-12}$~s~s$^{-1}$. Such properties imply particularly
strong surface dipole magnetic fields of the order of
$10^{14}-10^{15}$~G. About 24 magnetars with these properties are
known in our Galaxy, while one resides in the SMC and another in the
LMC. Most magnetars show high X-ray persistent luminosities, often
surpassing their rotational energy reservoir, hence, requiring an
extra source of power. The latter is believed to be of magnetic
origin, associated with their extremely strong outer and/or inner
magnetic fields.

Magnetars  are among the most variable sources within the NS
zoo. Almost all have been observed to emit short ($\sim0.1$~s), bright
($E_{\rm burst}\approx10^{37}-10^{41}$~erg), hard X-ray bursts
\citep[see][for reviews]{mereghetti15:mag,turolla15:mag}. Such 
bursting episodes can last days to weeks with varying number of bursts
emitted by a given source, ranging from 10s to 100s \citep[e.g.,
][]{israel08ApJ:1900,vanderhorst12ApJ:1550,lin11ApJ:1E1841}. These
bursting episodes are usually accompanied by changes in the source 
persistent X-ray emission; an increase by a factor of a few to a
$\sim$100 in flux level is usually observed to follow bursting episodes,
together with a hardening in the X-ray spectrum
\citep[e.g.,][]{kaspi14ApJ:1745,ng11ApJ:1547,esposito11MNRAS:1833,zelati15MNRAS:1745}.
Both properties usually relax quasi-exponentially to pre-burst levels
on timescales of weeks to months \citep{rea11:outburst}. Their pulse
properties also vary following bursting episodes, with a change in
shape and pulse fraction \citep[e.g.,][see
\citealt{woods06csxs:magnetars,mereghetti08AARv:magentars} for a
review]{woods04ApJ:1E2259,gogus02ApJ:1806}. We note that the
magnetar-defining observational characteristics mentioned above have
also been observed recently from NS not originally classified as
magnetars, i.e., the high-B pulsars PSR~J$1846-0258$
\citep{gavriil08Sci:psr1846} and PSR~J$1119-6127$
\citep{gogus16:j1119,archibald16:j1119}, the central compact object in
RCW~103 \citep{rea16:rcw103}, and a low-B candidate magnetar,
SGR~J$0418+5729$ \citep{rea13ApJ:0418}. Moreover, a surrounding wind
nebula, usually a pulsar associated phenomenon, has now been observed
from at least one magnetar, Swift~$1834.9-0846$
\citep{younes12ApJ:1834,younes16ApJ:1834,granot16:1834,torres17ApJ:1834}.

Magnetars also show bright hard X-ray emission ($>10$~keV) with total
energy occasionally exceeding that of their soft X-ray emission. This
hard emission is non-thermal in origin, phenomenologically described
as a power-law (PL) with a photon index ranging from $\Gamma\sim1-2$
\citep[see e.g., ][]{kuiper06ApJ}. The hard and soft component
properties may also differ \citep[e.g.,
][]{an13ApJ:1841,vogel14ApJ:2259,tendulkar15ApJ:0142}. In the context
of the magnetar model, the hard X-ray emission has been explained as
resonant Compton scattering of the soft (surface) emission by plasma
in the magnetosphere \citep{baring07,fernandez07ApJ,beloborodov13ApJ}.

So far, only four magnetars have been detected at radio frequencies,
excluding the high-B pulsar PSR~J$1119-6127$ that exhibited magnetar-like
activity \citep{weltevrede11MNRAS:1119,Antonopoulou2015MNRAS:1119,
  gogus16:j1119,archibald16:j1119}. The radio emission from the four
typical magnetars showed transient behavior, correlated with the X-ray
outburst onset \citep{camilo06Natur:1810,camilo07ApJ:1550,levin10ApJ}.
\citet{rea12ApJ:radiomag} showed that all radio magnetars have $L_{\rm
  X}/\dot{E}<1$ during quiescence. However, the physical mechanism for
the radio emission in magnetars (as well as why it has only been
detected in a very small number of sources)  remains largely unclear
\citep[e.g.,][]{szary15ApJ:magrad}, and could be inhibited if optimal
conditions for the production of pairs are not present
\citep[e.g.,][]{baring98ApJ}.

SGR~J$1935+2154$ is a recent addition to the magnetar family,
discovered with the \swift/X-Ray Telescope (XRT) on 2014 July 05
\citep{stamatikos14:1935}. Subsequent \swift, \chandra\ and \xmm\
observations taken in 2014 confirmed the source as a magnetar with a
spin period $P=3.25$~s and $\dot{P}=1.43\times10^{-11}$~s~s$^{-1}$, 
implying a surface dipole B-field of $B=2.2\times10^{14}$~G
\citep{israel16mnras:1935}. \src\ has been quite active since its
discovery with at least 3 other outbursts; 2015 February 22, 2016 May
14, and 2016 June 18. The source is close to the geometrical center of
the supernova remnant (SNR) G$57.2+0.8$ at a distance of $\sim9$~kpc
\citep{pavlovic13ApJS:SNR,sun11AA:SNR}.

In this paper, we report on the analysis of all X-ray observations of
\src\ taken after 2014 May, including a \nustar\ observation made
within days of the 2015 outburst identifying the broad-band X-ray spectrum
of the source. We also report on the analysis of radio observations
taken with Arecibo following the 2014 and 2016 June outbursts. X-ray
and radio observations and data reduction are reported in
Section~\ref{obs}, and analysis results are shown in
Section~\ref{res}. Section~\ref{discuss} discusses the results in the
context of the magnetar model, while Section~\ref{sum} summarizes our
findings.

\section{Observations and data reduction}
\label{obs}

\subsection{\chandra}

\chandra\ observed \src\ three times during its 2014 outburst, and
once during its 2016 June outburst. Two of the 2014 observations were
in Continuous-Clocking (CC) mode while the other two were taken in TE
mode with 1/8th sub-array. We analyzed these observations using CIAO
4.8.2, and calibration files CLADB version 4.7.2.

For the TE mode observations, we extracted source events from a circle
with radius 2\arcsec, while background events were extracted from an
annulus centered on the source with inner and outer radii of 4\arcsec\
and 10\arcsec, respectively. Source events from the CC-mode
observations were extracted using a box extraction region of 4\arcsec\
length. Background events were extracted from two box regions with the
same length on each side of the source region. We used the CIAO
\texttt{specextract}\footnote{http://cxc.harvard.edu/ciao/ahelp/specextract.html}
script to extract source and background spectral files, including
response RMF and ancillary ARF files. Finally, we grouped the spectra
to have only 5 counts per bin. Table~\ref{logObs} lists the details of
the \chandra\ observations.

\subsection{\xmm}

We analyzed all of the 2014 \xmm\ observations of \src. In all cases,
the EPIC-pn \citep{struder01aa} camera was operated in Full Frame
mode. The MOS cameras, on the other hand, were operated in small
window mode. Both cameras used the medium filter. All data products
were obtained from the \xmm\ Science Archive
(XSA)\footnote{http://xmm.esac.esa.int/xsa/index.shtml} and reduced  
using the Science Analysis System (SAS) version 14.0.0. 

The PN and MOS data were selected using event patterns 0--4 and 0--12,
respectively, during only good X-ray events (``FLAG$=$0''). We
inspected all observations for intervals of high background, e.g., due
to solar flares, and excluded those where the background level was
above 5\% of the source flux. The source X-ray flux was never high
enough to cause pile-up.

Source events for all observations were extracted from a circle with
center and radius obtained by running the task  {\sl eregionanalyse}
on the cleaned event files. This task calculates the optimum centroid
of the count distribution within a given source region, and the radius
of a circular extraction region that maximizes the source signal to
noise ratio. Background events were extracted from a source-free
annulus centered at the source with inner and outer radii of 60\arcsec
and 100\arcsec, respectively. We generated response matrix files using
the SAS task {\sl rmfgen}, while ancillary response files were
generated using the SAS task {\sl arfgen}. Again, we grouped the
spectra to have only 5 counts per bin. Table~\ref{logObs} lists the
details of the \xmm\ observations.

\subsection{\swift}
\label{swiftDatRed}

The \swift/XRT is a focusing CCD, sensitive to photons in the energy
range of $0.2-10$~keV \citep{burrows05SSRv:xrt}. XRT can operate in
two different modes. The photon counting (PC) mode, which results in a
2D image of the field-of-view (FOV) and a time-resolution of 2.5~s,
and the windowed timing (WT) mode, which results in a 1D image with a
time resolution of 1.766~ms.

We reduced all 2014, 2015, and 2016 XRT data using
\texttt{xrtpipeline} version 13.2, and performed the analysis using
HEASOFT  version 6.17. We extracted source events from a 30\arcsec
radius circle centered on the source, while background events were
extracted from an annulus centered at the same position as the source
with inner and outer radii of 50\arcsec\ and 100\arcsec,
respectively. Finally, we generated the ancillary files with
\texttt{xrtmkarf}, and used the responses matrices in \texttt {CALDB}
v014. The log of the XRT observations is listed in
Table~\ref{logObs}.

All observations that resulted in a source number of counts $>30$ were
included in the analysis individually. Observations with source
number counts $<30$ were merged with other observations which were
taken within a 2 day interval. Any individual or merged observation
that did not satisfy the 30 source number counts limit were  excluded
from the analysis. However, most of these lost intervals were
compensated with existing quasi-simulntaneous \chandra\ and \xmm\
observations.

\begin{longtable}{ccc} 
\hline
\hline
Telescope/Obs.~ID & Date & Net Exposure \\
                                 & (MJD) &   (ks)     \\
\hline
\multicolumn{3}{c}{2014} \\
\hline
\swift-XRT/00603488000&56843.40&3.37 \\
\swift-XRT/00603488001&56843.52&9.90 \\
\swift-XRT/00603488003&56845.25&3.93 \\
\swift-XRT/00603488004&56845.98&9.31 \\
\swift-XRT/00603488006&56846.66&3.67 \\
\swift-XRT/00603488007&56847.60&3.63 \\
\swift-XRT/00603488008$^a$&56851.52&5.33 \\
\swift-XRT/00603488009$^a$&56851.32&2.95 \\
\chandra/15874&56853.59&9.13 \\
\swift-XRT/00603488011&56858.00&2.95 \\
\chandra/15875&56866.03&75.1 \\
\chandra/17314&56900.03&29.0 \\
\xmm/0722412501&56926.95&16.9 \\
\xmm/0722412601&56928.20&17.8 \\
\xmm/0722412701&56934.36&16.1 \\
\xmm/0722412801&56946.11&8.61 \\
\xmm/0722412901&56954.15&6.53 \\
\xmm/0722413001&56957.95&12.4 \\
\xmm/0748390801&56976.16&9.83 \\
\hline
\multicolumn{3}{c}{2015} \\
\hline
\swift-XRT/00632158000  & 57075.51 & 7.33 \\
\swift-XRT/00632158001  & 57075.80 & 1.80 \\
\swift-XRT/00632158002  & 57076.52 & 5.91 \\
\swift-XRT/00033349014  & 57078.18 & 3.13 \\
\nustar/90001004002  & 57080.22 & 50.6 \\
\swift-XRT/00033349015  & 57080.24 & 5.94 \\
\swift-XRT/00033349016  & 57085.31 & 3.94 \\
\swift-XRT/00033349017  & 57092.55 & 3.91 \\
\swift-XRT/00033349018  & 57102.00 & 4.37 \\
\swift-XRT/00033349019$^a$  & 57127.16 & 1.97 \\
\swift-XRT/00033349020$^a$  & 57127.77 & 2.94 \\
\swift-XRT/00033349021$^a$  & 57128.56 & 2.66 \\
\swift-XRT/00033349022$^a$  & 57129.10 & 0.85 \\
\swift-XRT/00033349023$^a$  & 57134.35 & 1.37 \\
\swift-XRT/00033349024  & 57220.96 & 1.98 \\
\swift-XRT/00033349025  &57377.70 & 3.94 \\
\hline
\multicolumn{3}{c}{2016} \\
\hline
\swift-XRT/00686761000&57526.38& 1.67 \\
\swift-XRT/00686842000$^a$&57527.24& 0.84 \\
\swift-XRT/00033349026$^a$&57527.77& 2.96 \\
\swift-XRT/00687123000$^a$&57529.84& 1.21 \\
\swift-XRT/00687124000$^a$&57529.85& 0.81\\
\swift-XRT/00033349028$^a$&57539.87& 2.78\\
\swift-XRT/00033349029$^a$&57540.54& 0.47\\
\swift-XRT/00033349031&57554.16& 2.57 \\
\swift-XRT/00033349032&57561.02& 1.58 \\
\swift-XRT/00701182000&57562.81& 1.65 \\
\swift-XRT/00701590000&57565.58& 1.39 \\
\swift-XRT/00033349033$^a$&57567.18& 2.01 \\
\swift-XRT/00033349034$^a$&57569.52& 2.38 \\
\chandra/18884&57576.23& 18.2\\
\swift-XRT/00033349035&57576.77& 2.78\\
\swift-XRT/00033349036&57586.20& 2.48\\
\swift-XRT/00033349037&57597.04& 2.84\\
\hline
\caption{Log of X-ray observations}
\label{logObs}
\end{longtable}

\subsection{\nustar}

The Nuclear Spectroscopic Telescope Array (\nustar,
\citealt{harrison13ApJ:NuSTAR}) consists of two similar focal-plane
modules (FPMA and FPMB) operating in the energy range $3-79$~keV. It
is the first hard X-ray ($>10$~keV) focusing telescope in orbit.

\nustar\ observed \src\ on 2015 February 27 at 05:16:20 UTC. The net
exposure time of the observation is 50.6~ks (Table~\ref{logObs}).  We
processed the data using the \nustar\ Data Analysis Software,
\texttt{nustardas} version v1.5.1. We analyzed the data using the
\texttt{nuproducts} task (which allows for spectral extraction and
generation of ancillary and response files) and HEASOFT version
6.17. We extracted source events around the source position using a
circular region with 40\arcsec\ radius. Background events were
extracted from an annulus around the source position with inner and
outer radii of 80\arcsec\ and 160\arcsec, respectively.

\subsection{Arecibo Observations}

\label{subsec:DA}
\begin{table*}[th!]
\normalsize
\centering
\begin{minipage}{\textwidth}
\caption{\normalsize{Arecibo Observations Summary}}
\vspace{0.35 cm}
\centering
\scalebox{0.95}{
\begin{tabular}{|l|l|l|l|}
\hline
Ind & Project Id & \pbox{32cm}{Obs. Start Date} & \pbox{6cm}{Integration Time (hr)} \\ 
\hline
\hline
\multicolumn{4}{|c|}{\textbf{C-band observations}} \\
\hline
\multicolumn{4}{|c|}{$G = 8$\,K/Jy, $T_{sys} = 28$\,K}\\
\hline
\hline
1 & p2976 & 2015-03-05 & 1.0\\
2 & p2976 & 2015-03-12 & 1.0\\
3 & p2976 & 2015-03-27 & 1.3\\
4 & p3100 & 2016-07-05 & 0.7\\
5 & p3100 & 2016-07-12 & 1.0\\
6 & p3100 & 2016-07-27 & 0.3\\
\hline
\multicolumn{4}{|c|}{\textbf{L-band observations}} \\
\hline
\multicolumn{4}{|c|}{$G = 10$\,K/Jy, $T_{sys} = 33$\,K}\\
\hline
\hline
1 & p2976 & 2015-03-05  &1.0\\
2 & p2976 & 2015-03-12 & ---\\
3 & p2976 & 2015-03-27  &1.0\\
4 & p3100 & 2016-07-05  & 0.5\\
5 & p3100 & 2016-07-12 & ---\\
6 & p3100 & 2016-07-27  & 0.4 \\
\hline
\end{tabular}}
\label{table:obssum}
\end{minipage}
\vspace{0.5 in}
\end{table*}

We observed SGR~J1935+2154 with the 305-m William E. Gordon Telescope
at the Arecibo Observatory in Puerto Rico, as part of Director's
Discretionary Time, to search for radio emission after its X-ray
activation, both in 2015 and in 2016. The source was observed on 2015
March 5th, 12th and 27th (henceforth Obs. $1-3$) and on 2016 August
5th, 12th and 27th (Obs. $4-6$).  Observation durations ranged from
$\sim 1-2.5$hr; in each session (with the exception of Obs. 2 and 4)
the observation time was split between two different observing
frequencies.  A short summary of all observations is presented in
Table \ref{table:obssum}.

Observations using the Arecibo C-band receiver were performed at a
central frequency of $4.6$\,GHz, with the Mock Spectrometers as a
backend.  We used a bandwidth of $\sim172$\,MHz, which was split
across $32$ channels. The time resolution was $65~\mu$s with
$16$-bit samples. In every C-band observation we used the nearby
PSR~B$1919+21$ to test the instrumental setup.

The Arecibo L-band Wide receiver was used in the frequency range
$0.98-1.78$\,GHz with a central frequency of $1.38$\,GHz.  As backend,
we used the Puerto-Rican Ultimate Pulsar Processing Instrument
(PUPPI).  PUPPI provided $800$\,MHz bandwidth (roughly 500\,MHz
usable), split across $2048$ spectral channels. For our observations,
PUPPI was used in Incoherent Search mode. The data were sampled at
$40.96$\,$\mu$s with $8$ bits per sample.  At the start of every
L-band observation, PSR~J$1924+1631$ was observed to verify the
setup.

\section{Results}
\label{res}

\subsection{X-ray imaging}
\label{imres}

To assess the presence of any extended emission around \src\ we relied
on the four \chandra\ observations, as well as the 2014 \xmm\
observations. 


\begin{figure*}[]
\begin{center}
\includegraphics[angle=0,width=0.33\textwidth]{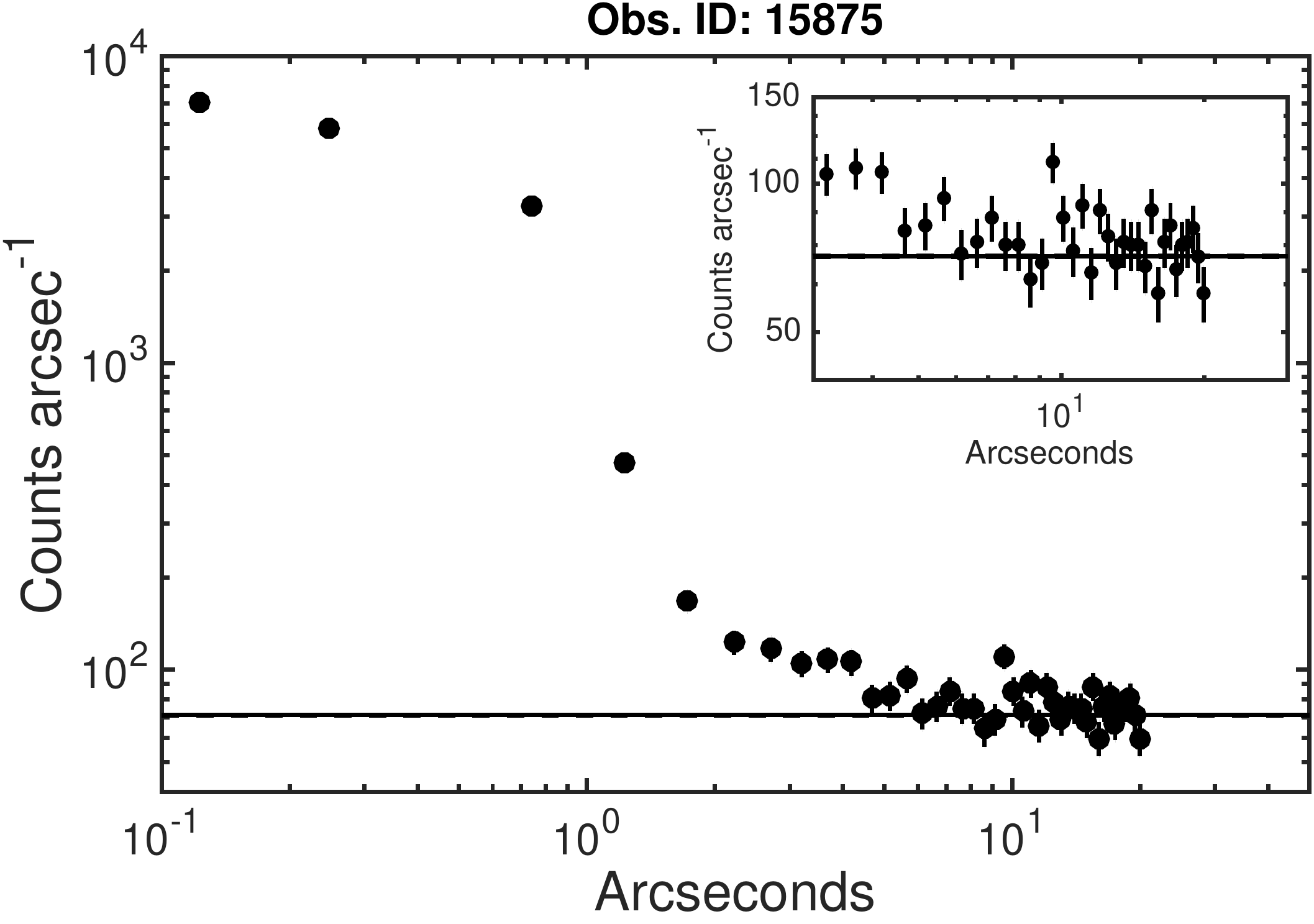}
\includegraphics[angle=0,width=0.33\textwidth]{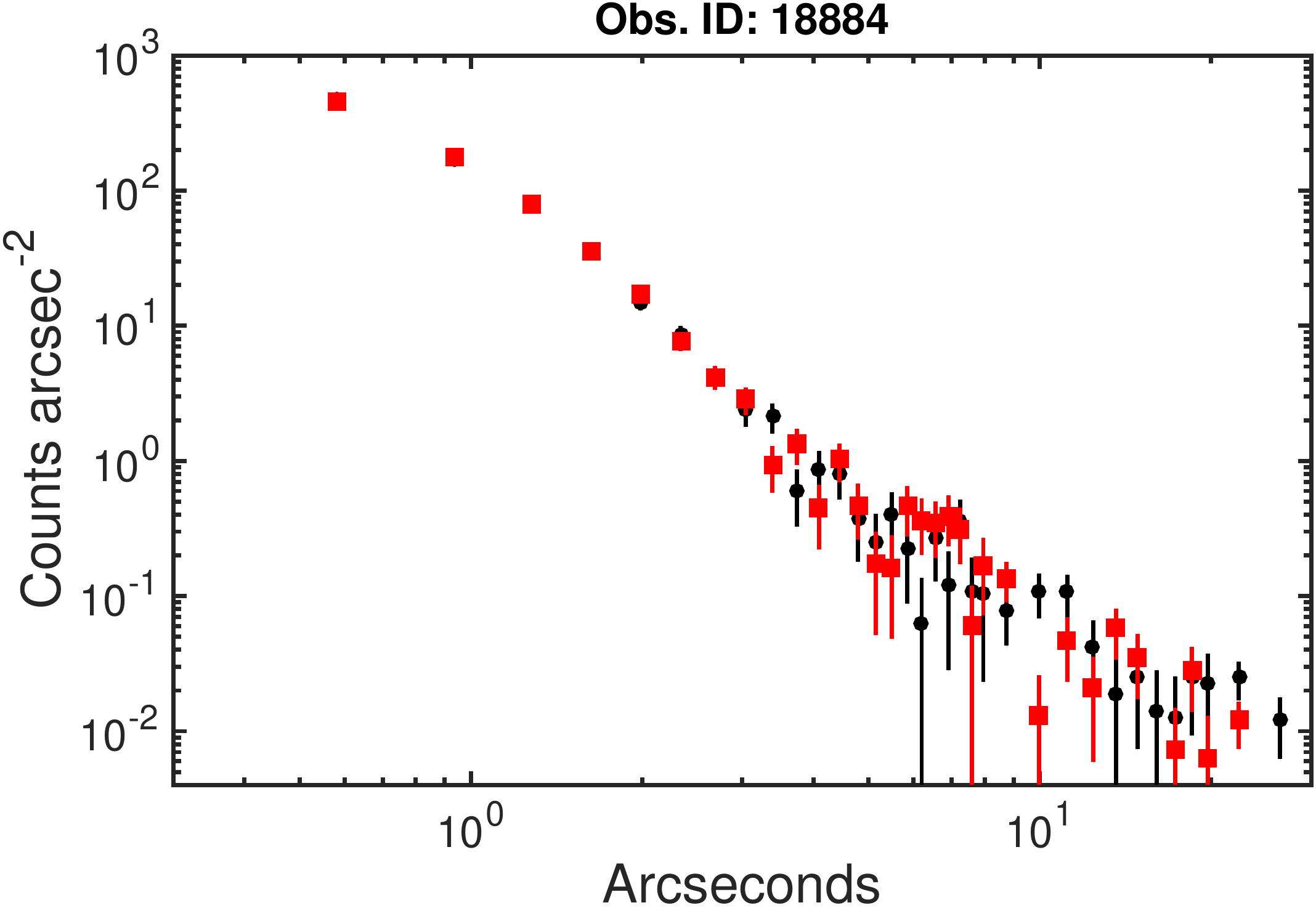}
\includegraphics[angle=0,width=0.33\textwidth]{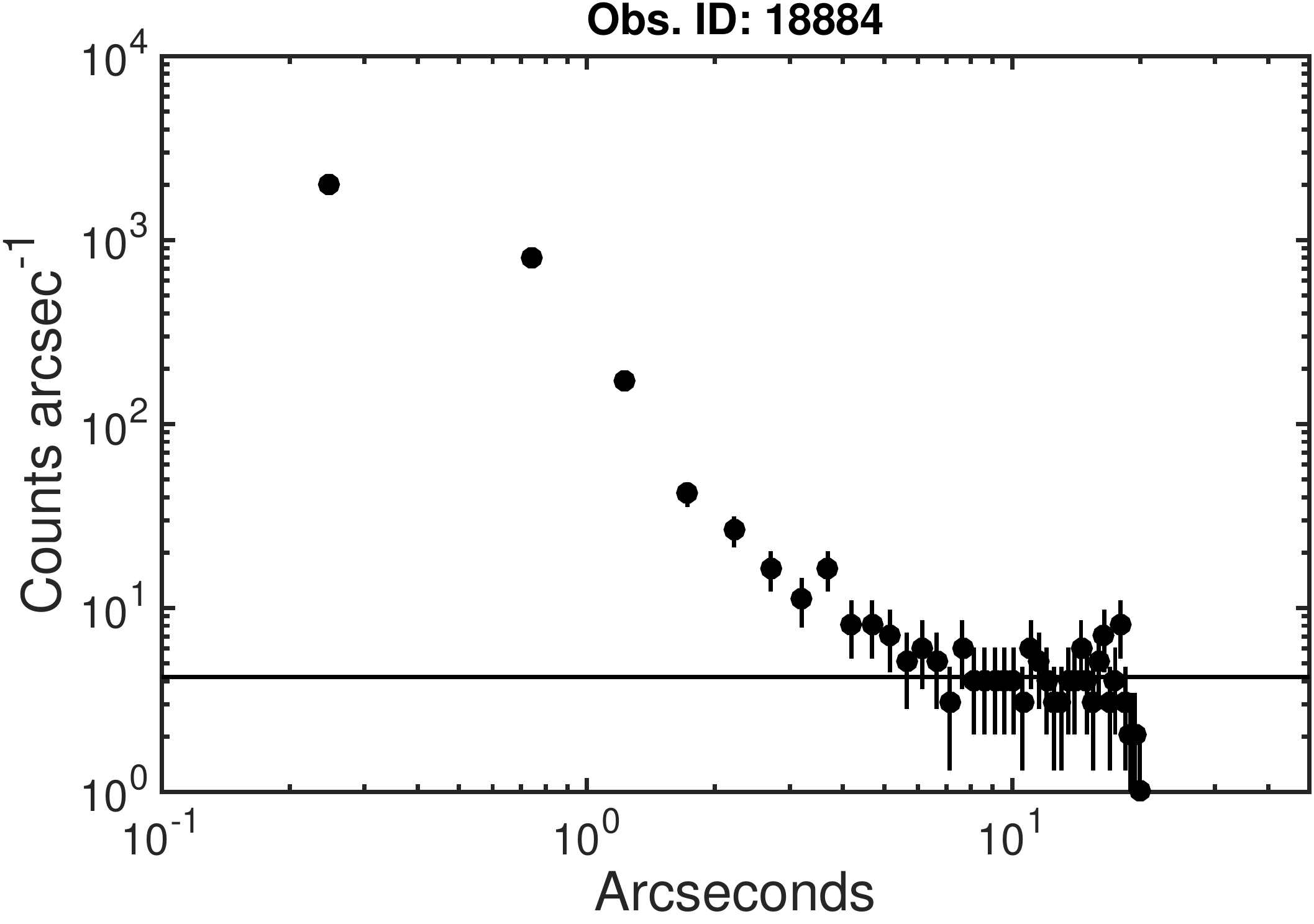}
\caption{{\sl Left panel.} \chandra\ 1-D radial profile from the \chandra\ 
  CC-mode observation 15875. The black horizontal line represents the
  background level, while black-dots represent the radial profile. No
  extended emission is obvious beyond 5\arcsec\ from the central
  brightest pixel. {\sl Middle panel.} \chandra\ 2-D radial profile
  from the 2016 TE mode observation. The black-dots represent the
  data radial profile, while the red squares represent the Chart and MARX
  PSF simulation. The agreement between the profile and the PSF
  simulation indicates the absence of any extended emission around
  \src. {\sl Right panel.} \chandra\ 1-D radial profile after
  converting the TE-mode observation 18884 into a 1D CC-mode
  observation, as a verification of our results for the CC-mode
  observation 15875. See text for more details.} 
\label{radProf}
\end{center}
\end{figure*}

Two of the \chandra\ observations, including the one in 2016, were
taken in TE mode, while the other two were taken in CC mode. For the
two TE mode observations we simulated a \chandra\ PSF at the source
position with the spectrum of \src, using the \chandra\ ray-trace
(ChaRT\footnote{http://cxc.harvard.edu/ciao/PSFs/chart2/}) and
MARX\footnote{http://space.mit.edu/CXC/MARX/}. The middle panel of
Figure~\ref{radProf} shows the radial profile of the 2016 TE mode
observations, which had an exposure twice as long as the one taken on
2014. Black dots represent the radial profile of the actual
observation, while the red squares represent the radial profile of the
simulated PSF. There is no evidence for small-scale extended emission
beyond a point source PSF in this observation. The 2014 observation
showed similar results.

The CC-mode observations are not straightforward to perform imaging
analysis with, given their 1-D nature. To mitigate this limitation, we
calculated and averaged the total number counts in each two pixels at
equal distance from the central brightest pixel, up to a distance of
20\arcsec\ (we also split the central brightest pixel into two, to
better sample the inner 0.5\arcsec). The background for these
observations was estimated by averaging the number of counts from all
pixels at a distance 25-50\arcsec\ from both sides of the central
brightest pixel. The left panel of Figure~\ref{radProf} shows the
results of our analysis on the longest of the 2 CC-mode observations,
obs. ID 15875 (notice the y-axis unit of counts~arcsec$^{-1}$). The
solid horizontal line represents the background level, while the dots
represent the 1-D radial profile of \src. The inset is a zoom-in at
the $3-20$\arcsec\ region. The level of emission beyond
$\sim5$\arcsec\ from the central pixel is consistent with the
background, hence, we conclude that there is no evidence for
small-scale extended emission from the source. We verified our results
by converting our 2016 TE-mode observation into a CC mode one by
collapsing the counts into 1D. We then performed the same analysis on
this converted image as the one done on the CC mode observations. The
results are shown in the right panel of Figure~\ref{radProf}. We note
that these results are in contrast with the results presented in
\citet[][see their Figure~1]{israel16mnras:1935}. We have not been
able to identify the reason of this discrepancy.

\xmm\ observations showed a weak extended emission after stacking all
seven 2014 observations, in accordance with the results reported by
\citet{israel16mnras:1935}.


\subsection{Timing}
\label{timAna}

\subsubsection{X-ray}

We searched the \nustar\ and \chandra\ data for the pulse period from
\src. We focused the search in an interval around the expected pulse
period from the source at the \nustar\ and \chandra\ MJDs, after
extrapolating the timing solution detected with \chandra\ and \xmm\
during the 2014 outburst \citep{israel16mnras:1935}. We included the
possibility of timing noise and/or glitches and searched an interval
with $\delta\dot{P}\approx1.0\times10^{-9}$~s~s$^{-1}$. For the FPMA
and FPMB modules, we extracted events from a circle with a 45\arcsec\
radius around the source position, and in the energy range
$3-50$~keV. We extracted the \chandra\ events using a 2\arcsec\ circle
centered at the source in the energy range $1-8$\,keV. We
barycenter-corrected the photon arrival times to the solar system
barycenter.

We first applied the $Z^2_m$ test algorithm \citep{buccheri83AApulse}
at the \nustar\ data, where $m$ is the number of harmonics. Although
the signal during the 2014 outburst was near-sinusoidal, we applied
the test using $m=1, 2, 3,$ and $5$, considering the possibility of a
change in the pulse shape during the later outbursts. The highest peak
in the $Z^2$ power in the \nustar\ data, with a significance of
$3.7\sigma$, is located at the period reported in
\citet{younes15:1935} of 3.24729(1)~s. This is largely different
compared to the pulse-period of 3.24528(6)~s derived by
\citet{israel16mnras:1935} for the 2015 \xmm\ observation taken a
month later. The change in frequency between the two observations is
about $1.2\times10^{-9}$~s~s$^{-1}$; too large to correspond to any
timing noise. We also repeated our above analysis for different energy
cuts, namely $3-10$~keV and $3-30$~keV, and for different circular
extraction regions of 30\arcsec\ and 37\arcsec\ radii (to optimize
S/N). We find no other significant peaks in the $Z^2$ power for any of
the above combinations. We, therefore, conclude that we do not detect
the spin period of the source in our 2015 \nustar\
observation. Following the same method, we searched for the pulse
period in the 2016 \chandra\ observation. Similarly, we do not detect
the pulse period from \src.

We estimated upper limits on the rms pulsed-fraction (PF) of a pure
sinusoidal modulation by simulating 10,000 light curves with mean
count rate corresponding to the true background-corrected count rate
of the source and pulsed at a given rms PF. For our 2015 \nustar\
observation, we derive a $3\sigma$ ($99.73\%$ confidence) upper-limit
on the rms PF of $26\%$, $35\%$, and $43\%$ in the energy
ranges $3-50$~keV, $3-10$~keV, and $10-50$~keV, respectively. For our
2016 \chandra\ observation, we set a $3\sigma$ rms PF upper-limit
of $8\%$. These limits are consistent with the $5\%$ rms pulsed
fraction derived during the 2015 \xmm\ observation in the 0.5-10~keV
range \citep{israel16mnras:1935}.

\subsubsection{Radio}

A consistent method of data analysis was adopted for both the C-band
and L-band data analysis, and was based on tools from the pulsar
search and analysis software PRESTO
\citep{SCOTT:2001,REM:2002,RCE:2003}.  To excise radio frequency 
interference (RFI) we created a mask using {\tt rfifind}.  After RFI
excision, we used two different techniques to search for radio
pulsations: i) a search based on the known spin parameters from an
X-ray derived ephemeris, and ii) a blind, Fourier-based periodicity
search, as we describe below.

{\sl Ephemeris based search.} Coherent X-ray pulsations from
SGR~J1935+2154 were detected by \cite{IRZ:2014} at a $> 10 \sigma$
confidence level. Upon this discovery, SGR~J1935+2154 was monitored
using \textit{XMM-Newton} and \textit{Chandra} observations between
$2014-2015$ \cite[see,][]{IER:2016}. This campaign resulted in a
timing solution as presented in Table 2 of \cite{IER:2016}. We used
the period, period derivative and second period derivative from this
ephemeris to extrapolate the source spin period for Obs. $1-3$ and
Obs. $4-6$ ($3.2452679$\,s, $3.2458097$\,s, respectively).  We then
folded each C-band and L-band observation with the appropriate spin
period using PRESTO's {\tt prepfold}.  This folding operation was
restricted to only optimize S/N over a search range in pulse period
and incorporated the RFI mask. We repeated this folding routine over
dispersion measures (DMs) ranging from $0$ to $1000$\,pc\,cm$^{-3}$ in
steps of $50$\,pc\,cm$^{-3}$.  Recently, using the intermediate flare
from SGR~J1935+2154 along with a magnetic field estimate from the
timing analysis of \cite{IER:2016}, \cite{KIS:2016} showed that the
magnetar is at a distance of $<10$\,kpc. We used the NE2001 Galactic
electron density model and integrated in the source direction up to
$10$\,kpc to obtain an expected DM. We obtain a value of $344$\,pc
$~$cm$^{-3}$ (typical error is 20\% fractional), which lies well
within the DM range of our searches. These searches found no plausible
radio pulsations from SGR~J1935+2154.

{\sl Blind searches.} We also conducted searches using no {\it a
  priori} assumption about the spin period in order to allow for a
change compared to the ephemeris (e.g., a glitch) or the serendipitous
discovery of a pulsar in the field.  Using {\tt prepsubband}, we
created barycentered and RFI-excised time series for a DM range of $0$
to $1050$\,pc\,cm$^{-3}$, where the trial DM spacing was determined
using {\tt ddplan}.  We then Fourier transformed each time series with
{\tt realfft} and conducted {\tt accelsearch} based searches (with a
maximum signal drift of $z_{\rm max} = 100$ in the power spectrum) in
order to maintain sensitivity to a possible binary orbit. The most
promising candidates from this search were collated and ranked using
{\tt ACCEL\_sift}.  We folded the raw filterbank data for the best 200
candidates identified with {\tt ACCEL\_sift} and then visually
inspected each candidate signal using parameters such as cumulative
S/N, S/N as a function of DM, pulse profile shape and broad-bandedness
as deciding factors in judging whether a certain candidate was
plausibly of astrophysical origin or whether it was likely to be noise
or RFI. We found no plausible astrophysical signals in this analysis
as well.

We estimate maximum flux density limits using the radiometer equation 
\citep[see][]{DTW:1985, Bhat:1998, LK:2012} given by: 

\begin{equation}
S_{min} = \frac{\left(\frac{S}{N}\right)~\beta~T_{sys}}{G~\sqrt{n_p~t_{obs}~\bigtriangleup f}}~\sqrt{\frac{W}{P-W}}
\end{equation}

where, $G$ is the gain of the telescope (K~Jy$^{-1}$), $\beta$ is a
correction factor which is $\sim 1$ for large number of bits per
sample, $T_{sys}$ is the system noise temperature (K), $\bigtriangleup
f$ is the bandwidth (MHz), and $t_{obs}$ is the integration time
(s) for a given source. These parameters for the observational setup
in each band are listed in Table\,\ref{table:obssum}.  We assume a
pulsar duty cycle ($W/P$) of $20\%$ and a minimum detectable
signal-to-noise ratio of $10$ in our search. This yields a maximum
flux density limit of $14$\,$\mu$Jy for the C-band observations and
$7$\,$\mu$Jy for the L-band observations.

\subsection{X-ray spectroscopy}
\label{specana}

We fit our spectra in the energy range $0.8-8$~keV for \chandra,
$0.8-10$~keV for \xmm\ and \swift, and $3-79$~keV for \nustar, using
XSPEC \citep{arnaud96conf} version 12.9.0k. We used the photoelectric
cross-sections of \citet{verner96ApJ:crossSect} and the abundances of
\citet{wilms00ApJ} to account for absorption by neutral gas. For all
spectral fits using different instruments, we added a multiplicative
constant normalization, frozen to 1 for the spectrum with the highest
signal to noise, and allowed to vary for the other instruments. This
takes into account any calibration uncertainties between the different
instruments. We find that this uncertainty is between $2-8\%$. For all
spectral fitting, we used the Cash-statistic in XSPEC for model
parameter estimation and error calculation, while the
\texttt{goodness} command was used for model comparison. All quoted
uncertainties are at the $1\sigma$ level, unless otherwise noted.

\subsubsection{The 2014 outburst}
\label{SecOut2014}

\begin{figure*}[t]
\begin{center}
\includegraphics[angle=0,width=0.49\textwidth]{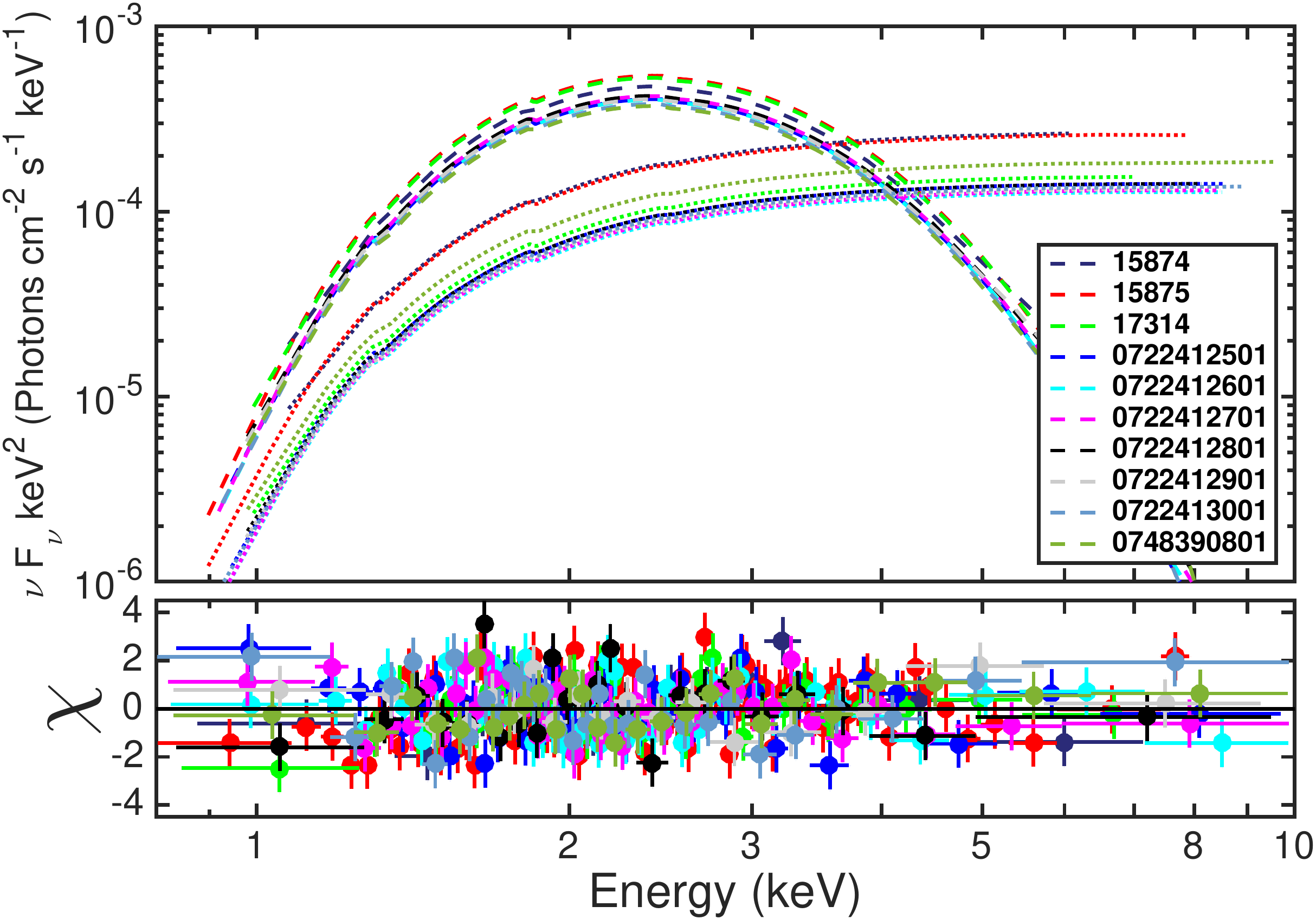}
\includegraphics[angle=0,width=0.49\textwidth]{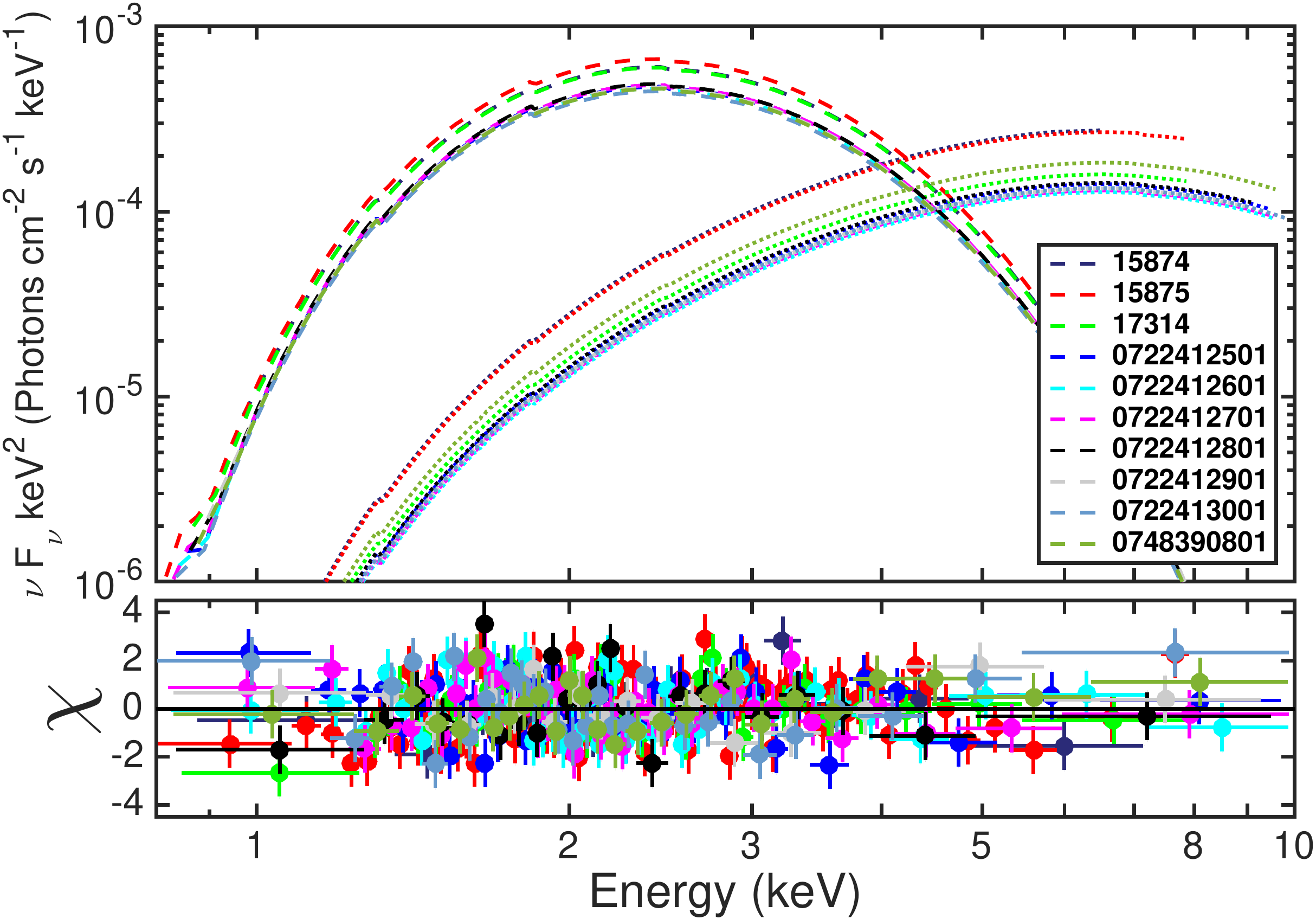}\\
\includegraphics[angle=0,width=0.49\textwidth]{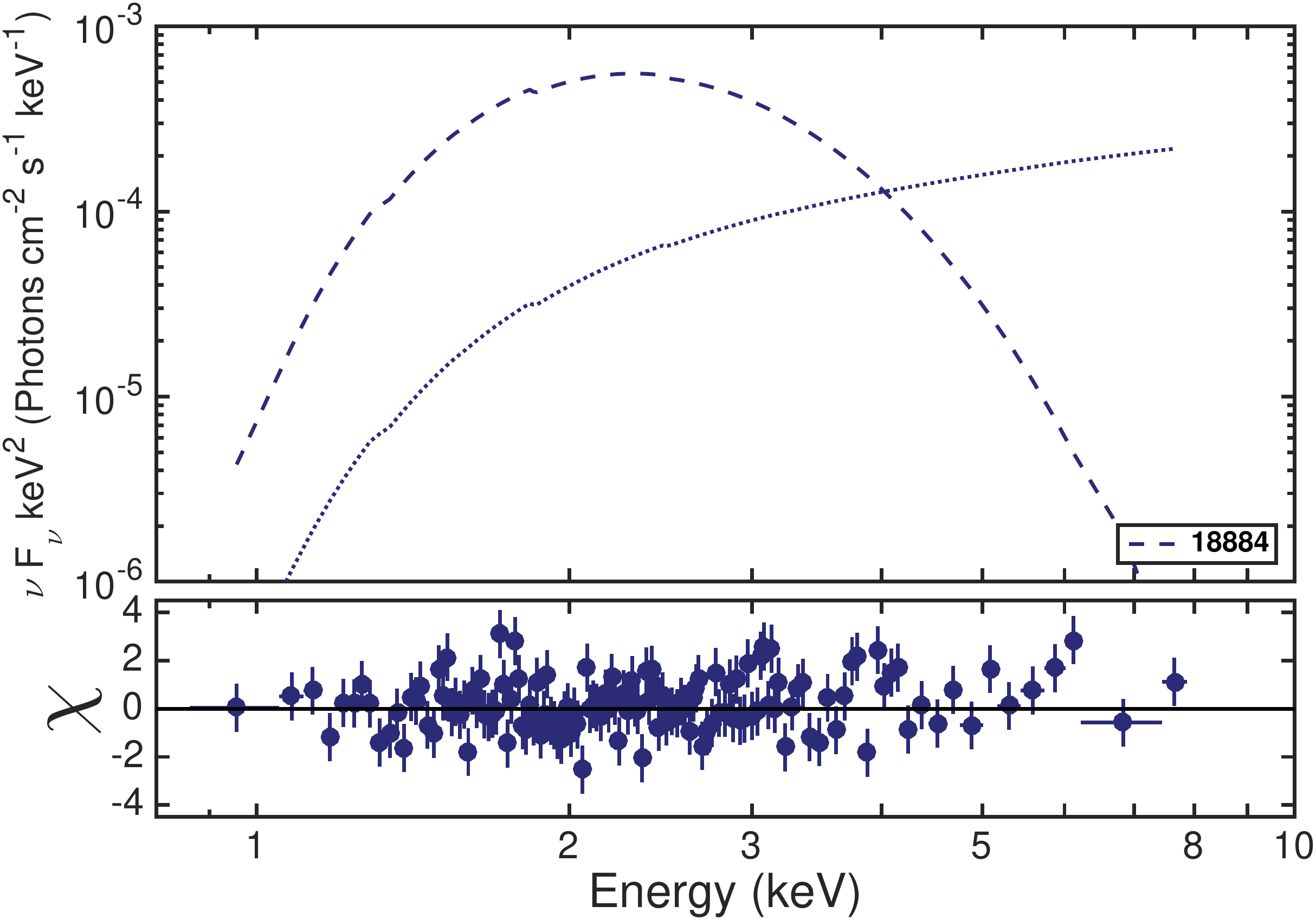}
\includegraphics[angle=0,width=0.49\textwidth]{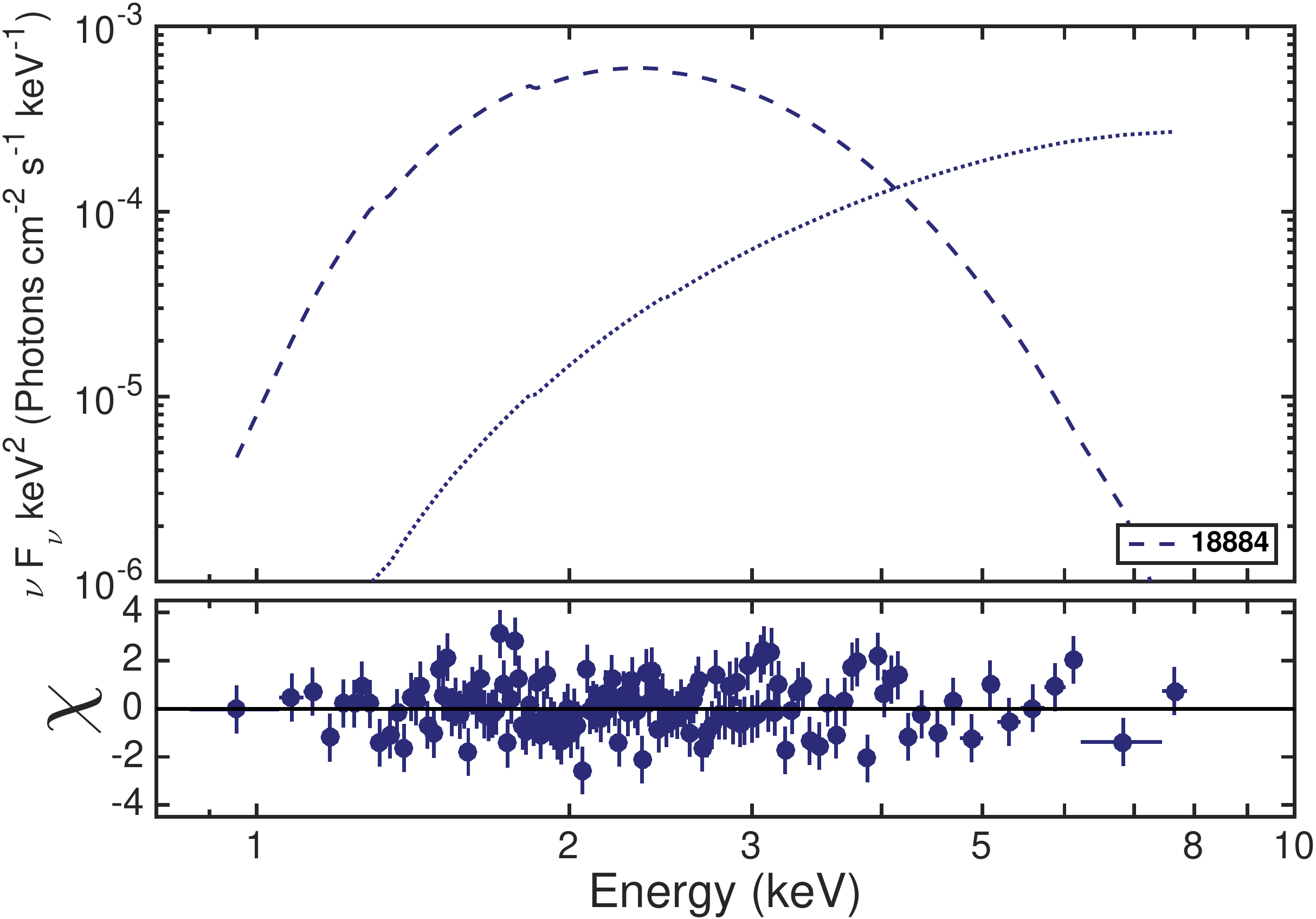}
\caption{{\sl Upper panels.} BB+PL ({\sl left}) and BB+BB ({\sl right}) fits
  to the \chandra\ and \xmm\ spectra from the 2014 outburst. {\sl
    Lower panels.} BB+PL ({\sl left}) and BB+BB ({\sl right}) fits
  to the \chandra\ spectrum from the 2016 outburst. The best-fit
  spectral components are shown in $\nu F_{\nu}$ space, while
  residuals are shown in terms of $\chi$. See text and
  table~\ref{specParam} for details.}
\label{specFit14}
\end{center}
\end{figure*}

We started our spectral analysis of the 2014 outburst
(Table~\ref{logObs}) by focusing on the high S/N ratio spectra derived
from the 7 \xmm\ observations (PN+MOS1+MOS2). Firstly, we fit these
spectra simultaneously with an absorbed (\texttt{tbabs} in XSPEC)
blackbody plus power-law (BB+PL) model, allowing all spectral model
parameters to vary freely, i.e., BB temperature (kT) and emitting
area, and PL photon index ($\Gamma$) and normalization, except for the
absorption hydrogen column density, which we linked between all
spectra. This model provides a good fit to the data with a C-stat of
5116.7 for 5196 degrees of freedom (d.o.f.). We find a hydrogen column
density $N_{\rm H} = (2.4\pm0.1)\times10^{22}$~cm$^{-2}$. The BB
temperatures and PL indices are consistent  for all spectra within the
$1\sigma$ level. Hence, we linked these parameters and re-fit. We find
a C-stat of 5131.75 for 5208 d.o.f. To estimate which model is
preferred by the data (here and elsewhere in the text), we estimate
the difference in the Bayesian Information Criterion (BIC), where
$\Delta BIC$ of 8 is considered significant and the model with the
lower $BIC$ is preferred \citep[e.g.,
][]{liddle07MNRAS:BIC}. Comparing the case of free versus linked $kT$
and $\Gamma$, we find that the case of linked parameters is preferred
with a $\Delta BIC\approx88$. This fit resulted in a hydrogen column
density $N_{\rm H} = 2.4\pm0.1\times10^{22}$~cm$^{-2}$, a BB
temperature $kT=0.46\pm0.01$~keV and area $R=1.45_{-0.03}^{+0.07}$~km,
and a photon index $\Gamma=2.0_{-0.5}^{+0.4}$.

\begin{table*}[th!]
\caption{Best-fit \xmm\ and \chandra\ X-ray spectral parameters}
\label{specParam}
\newcommand\T{\rule{0pt}{2.6ex}}
\newcommand\B{\rule[-1.2ex]{0pt}{0pt}}
\begin{center}{
\resizebox{1.0\textwidth}{!}{
\begin{tabular}{l c c c c c c c}
\hline
\hline
Obs. ID\T\B & $N_{\rm H}$ & $kT_{\rm cool}$ & $R_{\rm cool}^a$ & $\Gamma$/$kT_{\rm hot}$ & $R_{\rm hot}^a$ & $F_{\rm kT-cool}$ & $F_{\rm PL/kT-hot}$\\
\T\B & cm$^{-2}$ & (keV) & (km) & (/keV) &($10^{-3}$~km)& ($10^{-12}$,~erg s$^{-1}$ cm$^{-2}$) & ($10^{-12}$,~erg s$^{-1}$ cm$^{-2}$) \\
\hline
\multicolumn{8}{c}{2014 Outburst -- BB+PL}\\
\hline
15874 \T\B            & $2.46\pm0.08$ & $0.47\pm0.01$ & $1.7\pm0.08$ & $2.0\pm0.2$ & \ldots &  $1.78\pm0.16$ & $1.31\pm0.33$  \\
15875 \T\B            & (L)                      & (L)                      & $1.8\pm0.05$ & (L)                  & \ldots &  $2.01\pm0.09$ & $1.27\pm0.27$ \\
17314 \T\B            & (L)                      & (L)                      & $1.8\pm0.06$ & (L)                  & \ldots &  $1.96\pm0.08$ & $0.75\pm0.19$ \\
0722412501  \T\B & (L)                      & (L)                      & $1.6\pm0.05$ & (L)                  & \ldots &  $1.50\pm0.06$ & $0.69\pm0.17$ \\
0722412601  \T\B & (L)                      & (L)                      & $1.6\pm0.05$ & (L)                  & \ldots &  $1.49\pm0.06$ & $0.62\pm0.15$ \\
0722412701  \T\B & (L)                      & (L)                      & $1.6\pm0.05$ & (L)                  & \ldots &  $1.56\pm0.06$ & $0.64\pm0.16$ \\
0722412801  \T\B & (L)                      & (L)                      & $1.6\pm0.06$ & (L)                  & \ldots &  $1.57\pm0.07$ & $0.69\pm0.17$ \\
0722412901  \T\B & (L)                      & (L)                      & $1.6\pm0.06$ & (L)                  & \ldots &  $1.50\pm0.08$ & $0.65\pm0.17$ \\
0722413001  \T\B & (L)                      & (L)                      & $1.5\pm0.05$ & (L)                  & \ldots &  $1.42\pm0.07$ &  $0.66\pm0.17$ \\
0748390801  \T\B & (L)                      & (L)                      & $1.5\pm0.05$ & (L)                  & \ldots &  $1.38\pm0.09$ &  $0.90\pm0.21$ \\
\hline
\multicolumn{8}{c}{2016 Outburst}\\
\hline
18884  \T\B & $2.7\pm0.3$ & $0.42\pm0.04$ & $2.3\pm0.5$ & $1.3_{-0.7}^{+0.9}$ & \ldots & $2.0\pm0.3$ & $1.1\pm0.6$\\
\hline
\multicolumn{8}{c}{2014 Outburst -- BB+BB}\\
\hline
15874  \T\B & $2.30\pm0.04$ & $0.48\pm0.01$ & $1.8\pm0.6$ & $1.6\pm0.1$ & $80\pm9$ & $2.12\pm0.09$ & $0.53\pm0.08$\\
15875  \T\B                                         & (L) & (L) & $1.9\pm0.6$ & (L)                  & $79\pm9$ & $2.34\pm0.05$ & $0.52\pm0.03$\\
17314  \T\B                                         & (L) & (L) & $1.8\pm0.6$ & (L)                  & $61\pm8$ & $2.10\pm0.06$ & $0.31\pm0.04$\\
0722412501  \T\B                               & (L) & (L) & $1.6\pm0.5$ & (L)                  & $57\pm7$ & $1.66\pm0.04$ & $0.27\pm0.02$\\
0722412601  \T\B                               & (L) & (L) & $1.5\pm0.6$ & (L)                  & $54\pm7$ & $1.62\pm0.04$ & $0.25\pm0.02$\\
0722412701  \T\B                               & (L) & (L) & $1.6\pm0.5$ & (L)                  & $55\pm7$ & $1.70\pm0.04$ & $0.25\pm0.02$\\
0722412801  \T\B                               & (L) & (L) & $1.6\pm0.5$ & (L)                  & $57\pm7$ & $1.72\pm0.05$ & $0.28\pm0.03$\\
0722412901  \T\B                               & (L) & (L) & $1.6\pm0.6$ & (L)                  & $55\pm8$ & $1.65\pm0.05$ & $0.26\pm0.03$\\
0722413001  \T\B                               & (L) & (L) & $1.5\pm0.6$ & (L)                  & $56\pm7$ & $1.57\pm0.04$ & $0.26\pm0.03$\\
0748390801  \T\B                               & (L) & (L) & $1.5\pm0.6$ & (L)                  & $65\pm8$ & $1.62\pm0.05$ & $0.36\pm0.03$\\
\hline
\multicolumn{8}{c}{2016 Outburst}\\
\hline
18884  \T\B & $2.7\pm0.3$ &  $0.43\pm0.02$ & $2.3\pm0.4$ & $2.0_{-0.5}^{+1.3}$ & $52_{-26}^{+36}$ & $2.5\pm0.3$ & $0.45_{-0.06}^{-0.07}$\\
\hline
\end{tabular}}}
\begin{list}{}{}
\item[{\bf Notes.}] Fluxes are derived in the energy range 0.5-10~keV. $^a$ Assuming a distance of 9~kpc.
\end{list}
\end{center}
\end{table*}

We then fit the three \chandra\ spectra simultaneously, linking the
hydrogen column density, while leaving all other fit parameters free
to vary. We find a common hydrogen column density $N_{\rm
  H}=(2.9\pm0.3)\times10^{22}$~cm$^{-2}$. Similar to the case of the
\xmm\ observations, the BB temperature and the PL photon index were
consistent within the 1$\sigma$ confidence level among the three
observations. We, therefore, linked the BB temperature and the PL
photon index in the three observations and found $kT=0.46\pm0.02$,
$R=1.8\pm0.2$~km, and $\Gamma=2.4_{-0.6}^{+0.4}$.

\begin{figure*}[!h]
\begin{center}
\includegraphics[angle=90,width=0.5\textheight]{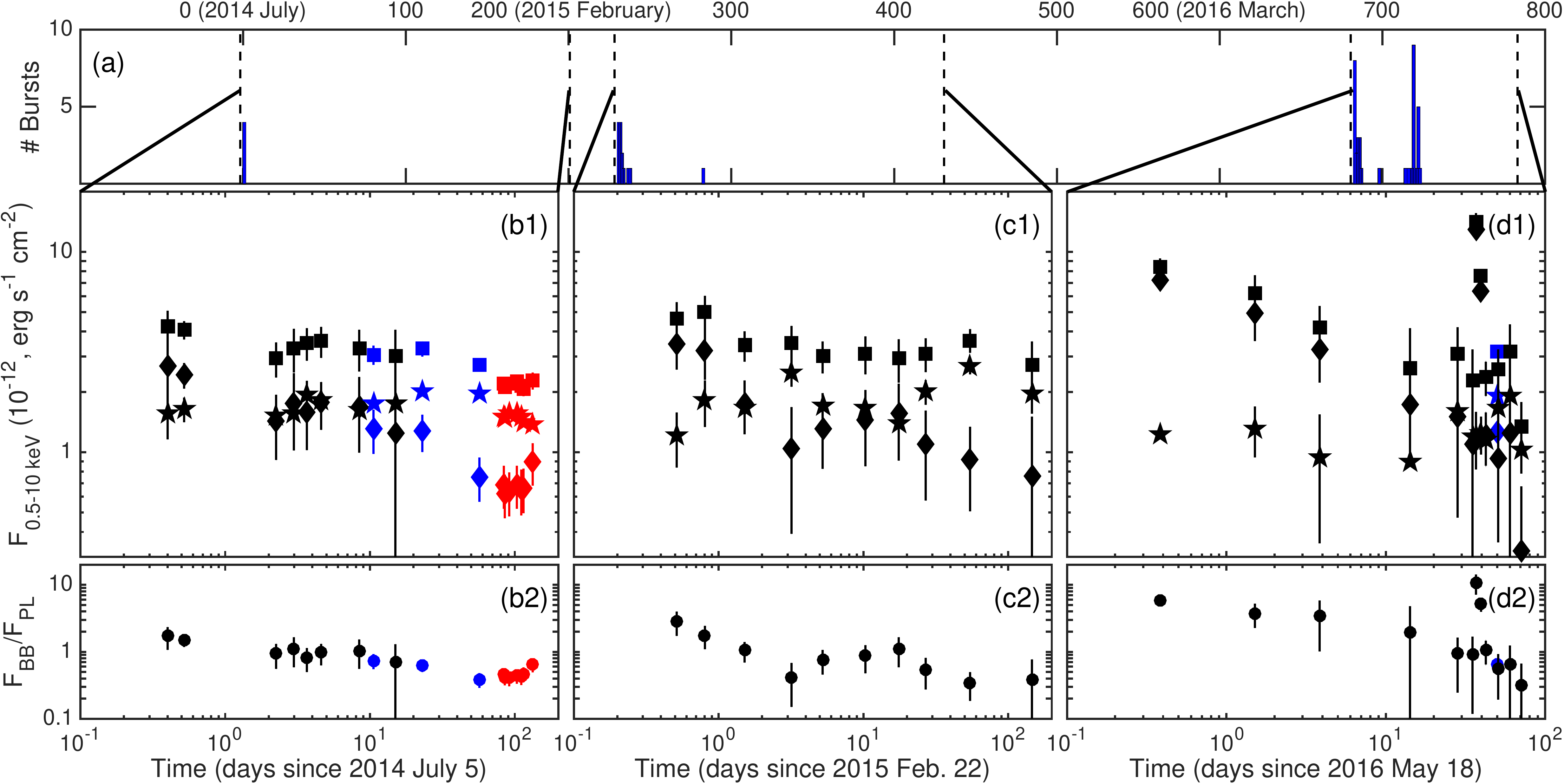}
\caption{\src\ BB+PL spectral evolution during the 2014, 2015, and May
  and June 2016 outbursts. {\sl Panel (a)} shows the number of bursts
  detected by the Inter Planetary Network (IPN) since the source discovery
  and up to August 2016. {\sl Panels (b1)} , {\sl (c1)}, and {\sl (d1)},
  represent the evolution of the BB (stars), PL (diamonds), and total
  fluxes (squares) from outburst onset and up to 200 days. {\sl
    Panels (b2)} , {\sl (c2)}, and {\sl (d2)}, represent the evolution
  of the $F_{\rm PL}/F_{\rm BB}$ ratio. Colors represent fluxes
  derived from different instruments (black:\swift, blue:\chandra,
  red:\xmm). See text for details.}
\label{specEvolPL}
\end{center}
\end{figure*}

\begin{figure*}[!h]
\begin{center}
\includegraphics[angle=90,width=0.5\textheight]{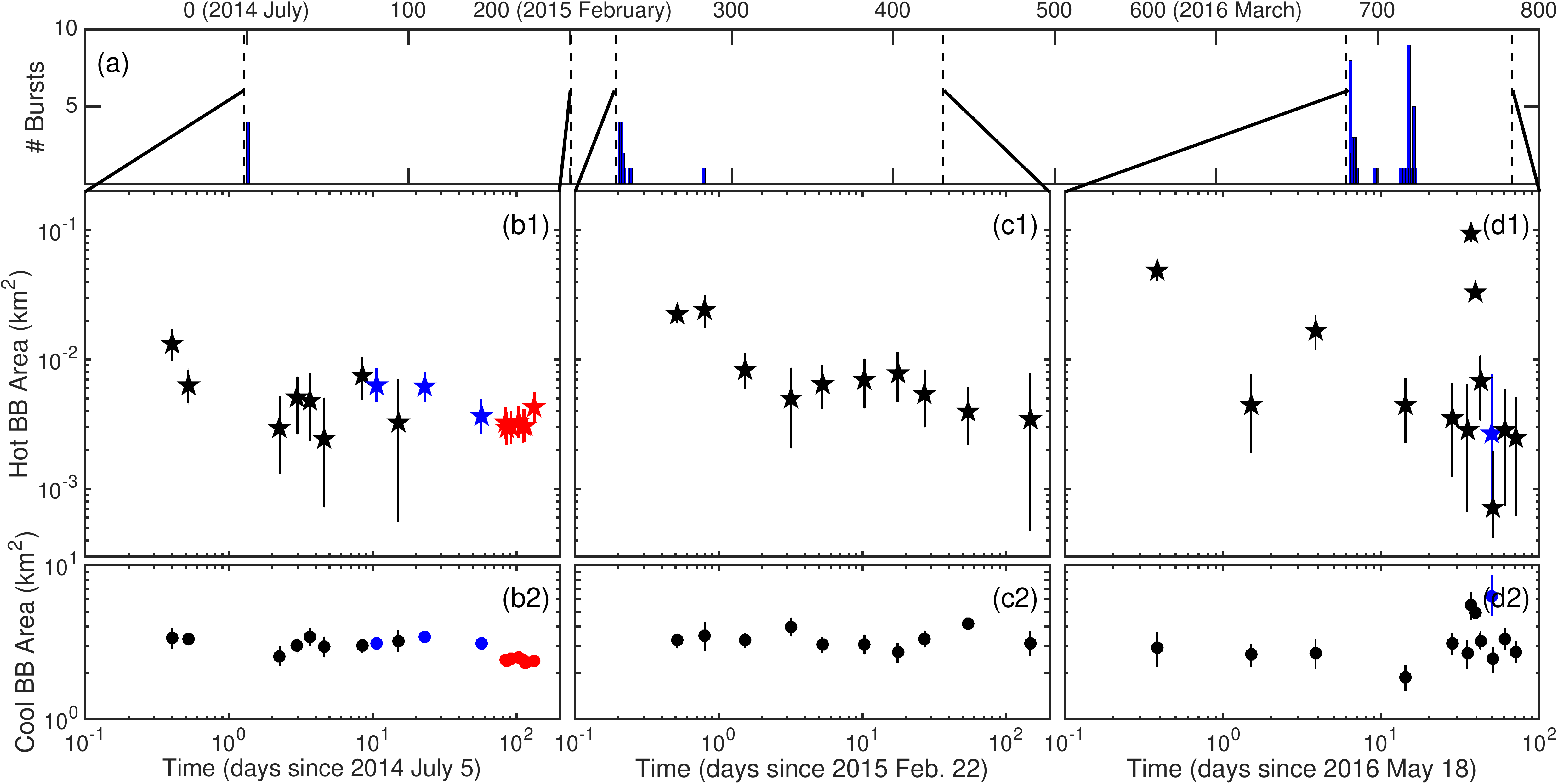}
\caption{\src\ BB+BB spectral evolution during the 2014, 2015, and May
  and June 2016 outbursts. {\sl Panel (a)} shows the number of bursts
  detected by the Inter Planetary Network (IPN) since the discovery
  and up to August 2016. {\sl Panels (b1)} , {\sl (c1)}, and {\sl (d1)},
  represent the evolution of the hot BB area from outburst onset and
  up to 200 days. {\sl Panels (b2)} , {\sl (c2)}, and {\sl
    (d2)}, represent the evolution of the cool BB area. Colors
  represent values derived from different instruments (black:\swift,
  blue:\chandra, red:\xmm). See text for more details.}
\label{specEvol2BB}
\end{center}
\end{figure*}

Given the consistency in $N_{\rm H}$, BB temperature and PL photon
index between the \chandra\ and \xmm\ observations, we then fit the
spectra from all 10 observations simultaneously, first, only linking
the $N_{\rm H}$ among all observations. We find a good fit with a
C-stat of 5750 for 5845 d.o.f, with
$N_{\rm}=(2.4\pm0.1)\times10^{22}$~cm$^{-2}$. Similar to the above two
cases, we find that the BB temperatures and the PL indices are
consistent within 1$\sigma$. Hence, we fit all 10 observations while
linking $KT$ and $\Gamma$. We find a C-stat of 5806 for 5863
d.o.f. Comparing this fit to the above case, we find a $\Delta
BIC$=100, suggesting that the latter fit is preferred over the fit
where parameters were left free to vary. The best fit spectral
parameters for the BB+PL model are summarized in
Table~\ref{specParam}, while the data and best fit model are shown in
Figure~\ref{specFit14}.

We also fit all spectra with an absorbed BB+BB model following the
above methodology. We first only link $N_{\rm H}$ among all spectra
while allowing the temperature and emitting area of the 2 BBs free to
vary. We find that the BB temperature of the cool component as well as
the hot component are consistent at the $1\sigma$ among all 10
observations, and were, therefore, linked. This alternative fit
resulted in a C-stat of 5812 for 5863 d.o.f, similar in goodness to
the BB+PL fit. Table~\ref{specParam} gives the BB+BB best fit spectral
parameters while the data and best fit model are shown in
Figure~\ref{specFit14}.

We analyzed the \swift/XRT observations taken during the 2014
outbursts following the procedure explained in
Section~\ref{swiftDatRed}. We fit all XRT spectra simultaneously with
the BB+PL and BB+BB models. Due to the limited statistics, we fixed
the temperatures and the photon indices to the values derived with the
above \xmm+\chandra\ fits. We made sure that the resulting fit was
statistically acceptable  using the XSPEC \texttt{goodness}
command. In the event of a statistically bad fit, we allowed the
temperatures and the photon indices to vary within the $3\sigma$
uncertainty of the \xmm+\chandra\ fits, which did give a statistically
acceptable fit in all cases. We show in Figure~\ref{specEvolPL} the
flux evolution of the BB+PL model and in Figure~\ref{specEvol2BB} the
areas evolution of the 2BB model. These results are discussed in
Section~\ref{discuss}.

Finally, we note that during the 2015 outburst, which will be
discussed in Section~\ref{SecOut2015}, \nustar\ reveals a hard X-ray
component dominating the spectrum at energies $>10$~keV and with a
non-negligible contribution at energies $5-10$~keV. In order to
understand the effect of such a hard component on the spectral shape
below $10$~keV (if it indeed exists during the 2014 outburst), we
added a hard PL component to the two above models (i.e., BB+PL and
2BB) while fitting the 7 \xmm\ observations. We fixed its index and
normalization to the result of a PL fit to the \nustar\ data from
$10-79$~keV\footnote{We used a simultaneous \swift/XRT observation to
  properly normalize the flux of this hard PL component to the 2014
  \xmm\ ones, assuming that the PL flux below and above 10~keV varies
  in tandem.}. As one would expect, we find that the addition of this
extra hard PL results in a softening of the $<10$~keV PL and hot BB
components. On average, we find a photon index for the soft PL
$\Gamma=2.7\pm0.3$. For the 2BB model, we find a temperature for the
hot BB $kT=0.8\pm0.2$ with a radius for the emitting area
$R\approx210\pm30$~m. Moreover, we find the fluxes of the low energy
PL or the hot BB to be a factor of $\sim$3 lower, however, the total
$0.5-10$~keV flux is similar to the above 2 models when we did not
include contribution from a hard PL. We cannot, unfortunately, add a
hard PL component to the XRT spectra and still extract meaningful flux
values from the 2 other components $<10$~keV, due to very limited
statistics. A complete statistical analysis, invoking many spectral
simulations, aiming at understanding the exact effect of a hard PL
component to the spectral curvature $<10$~keV is beyond the scope of
this paper. In all our discussions in Section~\ref{discuss}, however,
we made sure to avoid making any conclusions that could be affected by
such a shortcoming of the data we are considering here.

\subsubsection{The 2015 outburst}
\label{SecOut2015}


For the 2015 outburst, we first concentrated on the analysis of the
simultaneous \nustar\ and \swift/XRT observations (Table~\ref{logObs})
taken on February 27, 5 days following the outburst onset. This
provided the first look at the broad-band X-ray spectrum of the
source. \src\ is clearly detected in the two \nustar\ modules with a
background-corrected number of counts of $\sim800$ ($3-79$\,keV). We
find a background-corrected number of counts in the $3-10$~keV and
$10-79$~keV of about $500$ and $300$ counts, respectively. The
simultaneous XRT observation provided about 130 background-corrected
counts in the energy range $0.5-10$~keV. 

\begin{figure}[]
\begin{center}
\includegraphics[angle=0,width=0.49\textwidth]{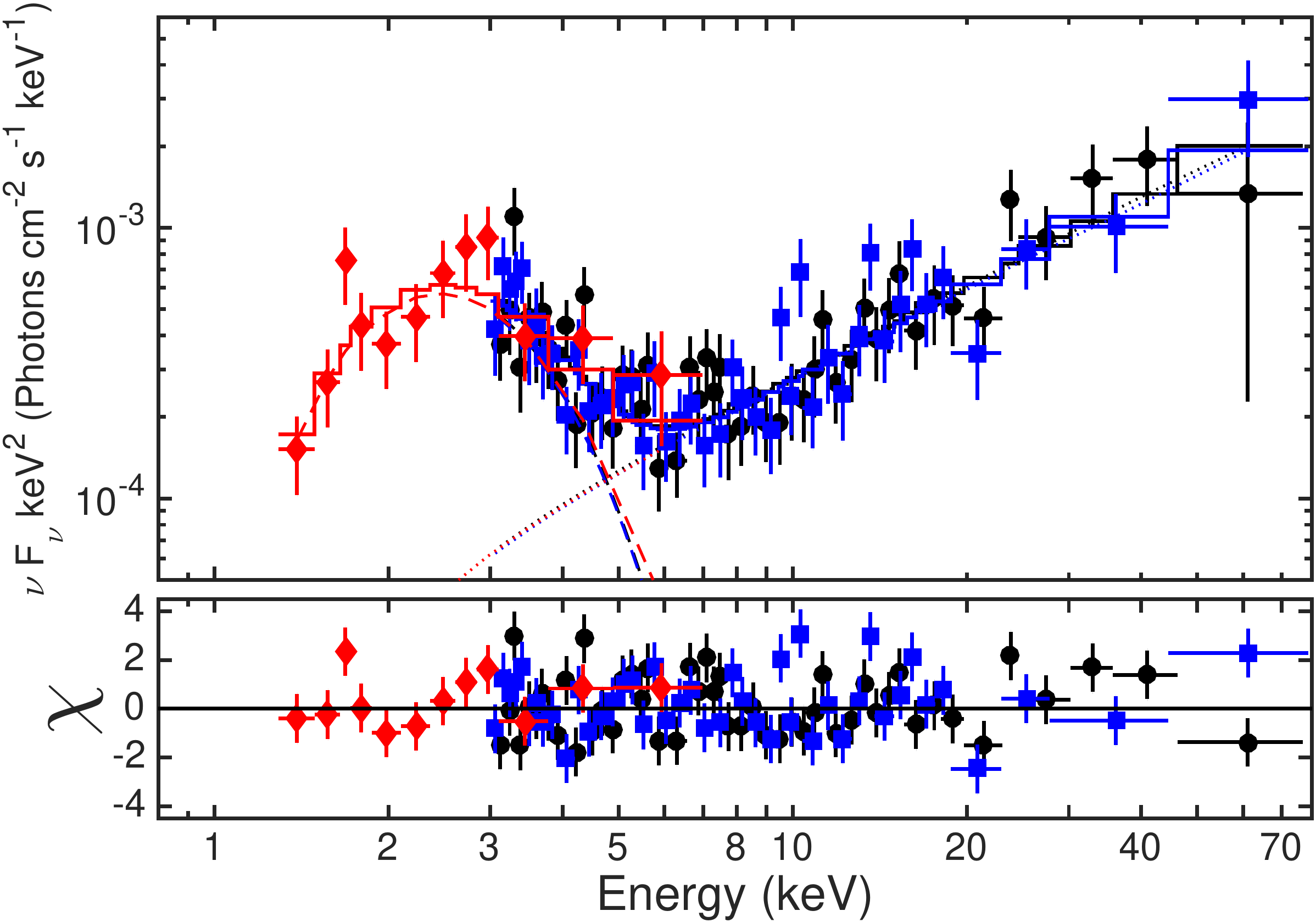}
\caption{{\sl Upper panel.} Simultaneous broad-band \nustar\ and
  \swift-XRT spectra of \src\ taken on 2015 February 27, 5 days after
  the 2015 outburst onset. Dots, squares, and diamonds are the
  \nustar\ FPMA, FPMB, and \swift-XRT spectra, respectively. The solid
  lines represent the absorbed BB+PL best fit model in $\nu F\nu$
  space, while the dashed and dotted lines represent the BB and PL
  components, respectively. {\sl Lower panel.} Residuals of the best
  fit are shown in terms of standard deviation.}
\label{specFit}
\end{center}
\end{figure}

We then fit the spectra simultaneously to an absorbed BB+PL model. We
find a good fit with a C-stat of 444 for 452 d.o.f., with an $N_{\rm
  H}=(2\pm0.7)\times10^{22}$~cm$^{-2}$.  We find a BB temperature
$kT=0.51\pm0.04$, a BB emitting area radius $R=1.4\pm0.3$~km, and a PL
photon index $\Gamma=0.9\pm0.1$. This spectral fit results in a
$0.5-10$~keV and $10-79$~keV absorption corrected fluxes of
$(2.6\pm0.4)\times10^{-12}$~erg~s$^{-1}$~cm$^{-2}$ 
and $(3.7\pm0.4)\times10^{-12}$~erg~s$^{-1}$~cm$^{-2}$, respectively.
Table~\ref{specParamNus} summarizes the best fit model parameters
while Figure~\ref{specFit} shows the data and best fit model
components in $\nu F\nu$ space ({\sl upper-panel}) and the residuals
in terms of $\sigma$ ({\sl lower-panel}).

Since the \chandra\ and \xmm\ 2014 spectra were best fit with a
2-component model below $10$~keV, we added a third component to the
\swift+\nustar\ data, a BB or a PL. Such a three model component is
required for many bright magnetars to fit the broad-band $0.5-79$~keV
spectra \citep[e.g., ][]{hascoet14ApJ}. For \src, the addition of
either component does not significantly improve the quality of the
fit, both resulting in a C-stat of 441 for 450 d.o.f. To understand
whether our \swift+\nustar\ data are of high enough S/N to exclude the
possibility of a 3 model component, we simulated 10,000 \swift-XRT and
\nustar\ spectra with their true exposure times, based on the 2014
$0.5-10$~keV spectrum and including a hard PL component as measured
above. We find that we cannot retrieve all three components at the
$3\sigma$ level; most of these simulated spectra are best fit with a 2
model component, namely an absorbed PL+BB. We, hence, conclude that
our \swift+\nustar\ data do not require the existence of a 3 component
model for the broad-band spectrum of \src.


\begin{table}[th!]
\caption{Best-fit spectral parameters to the 2015 simultaneous \swift-XTR and \nustar\ spectra.}
\label{specParamNus}
\newcommand\T{\rule{0pt}{2.6ex}}
\newcommand\B{\rule[-1.2ex]{0pt}{0pt}}
\begin{center}{
\begin{tabular}{l c}
\hline
\hline
\multicolumn{2}{c}{BB+PL}\\
\hline
$N_{\rm H}$ ($10^{22}$~cm$^{-2}$) \T\B & $2_{-0.7}^{+0.8}$\\
$kT$ (keV) \T\B & $0.51\pm0.04$ \\
$R_{\rm cool}^a$ (km) \T\B & $1.4_{-0.3}^{+0.5}$ \\
$F_{\rm BB}$ ($10^{-12}$~erg~s$^{-1}$~cm$^{-2}$) \T\B & $2.0_{-0.4}^{+0.5}$ \\
$\Gamma$ \T\B & $0.9\pm0.1$\\
$F_{\rm PL}$ ($10^{-12}$~erg~s$^{-1}$~cm$^{-2}$) \T\B & $4.2_{-0.7}^{+0.8}$ \\
\hline
$F_{\rm 0.5-10~keV}$ ($10^{-12}$~erg~s$^{-1}$~cm$^{-2}$) \T\B & $2.6\pm0.4$\\
$F_{\rm 10-79~keV}$ ($10^{-12}$~erg~s$^{-1}$~cm$^{-2}$) \T\B & $3.7_{-0.5}^{+0.4}$ \\
$L_{\rm 0.5-79 keV}^a$ ($10^{34}$~erg~s$^{-1}$) \T\B & $6.1\pm0.4$ \\
\hline
\end{tabular}}
\begin{list}{}{}
\item[{\bf Notes.}] $^a$Assuming a distance of 9~kpc.
\end{list}
\end{center}
\end{table}

To study the spectral evolution of the source during its 2015
outburst, we fit the \swift/XRT spectra of observations taken after
2015 February 22 (Table~\ref{logObs}) with an absorbed BB+PL and a 2BB
models. We fixed the absorption column density, temperatures and the
photon index to the values derived with the 2014 \xmm+\chandra\ fits,
but allowed for them to vary within their $3\sigma$ uncertainties in
the case of a statistically bad fit. We show in
Figure~\ref{specEvolPL} the flux evolution of the BB+PL model and in
Figure~\ref{specEvol2BB} the areas evolution of the 2BB model. These 
results are discussed in Section~\ref{discuss}.

\subsubsection{The 2016 outburst}
\label{SecOut2016}

We started our spectral analysis of the 2016 outburst with the
\chandra\ observation taken on July 07. Similar to the high S/N
spectra from the 2014 and 2015 outbursts, an absorbed BB or PL
spectral model fails to describe the data adequately. Hence, we fit 
an absorbed BB+PL and a 2BB model to the data. Both models result in
equally good fits with a C-stat of 289 for 302 d.o.f. The best fit
model parameters are shown in Table~\ref{specParam}, while the models
in $\nu F_{\nu}$ space and deviations of the data from the model in
terms of $\sigma$ are shown in Figure~\ref{specFit}. These spectral 
parameters are within $1\sigma$ uncertainty from the parameters
derived during the 2014 and 2015 outbursts. 

\src\ was observed regularly after the May outburst of 2016 with
\swift. These observations also covered its 2016 June outburst. We
analyzed all XRT observations taken during this period, and fit all
spectra with an absorbed BB+PL and 2BB models. We froze the absorption
hydrogen column density $N_{\rm H}$, $\Gamma$ and $kT$ to the best fit
values as derived during the 2014 outburst. The evolution of the flux
for the BB+PL model and the emitting area radius of the 2BB model are
shown in Figures~\ref{specEvolPL} and \ref{specEvol2BB},
respectively. 

\subsection{Outburst comparison and evolution}
\label{outComapre}

\begin{figure}[th]
\begin{center}
\includegraphics[angle=0,width=0.49\textwidth]{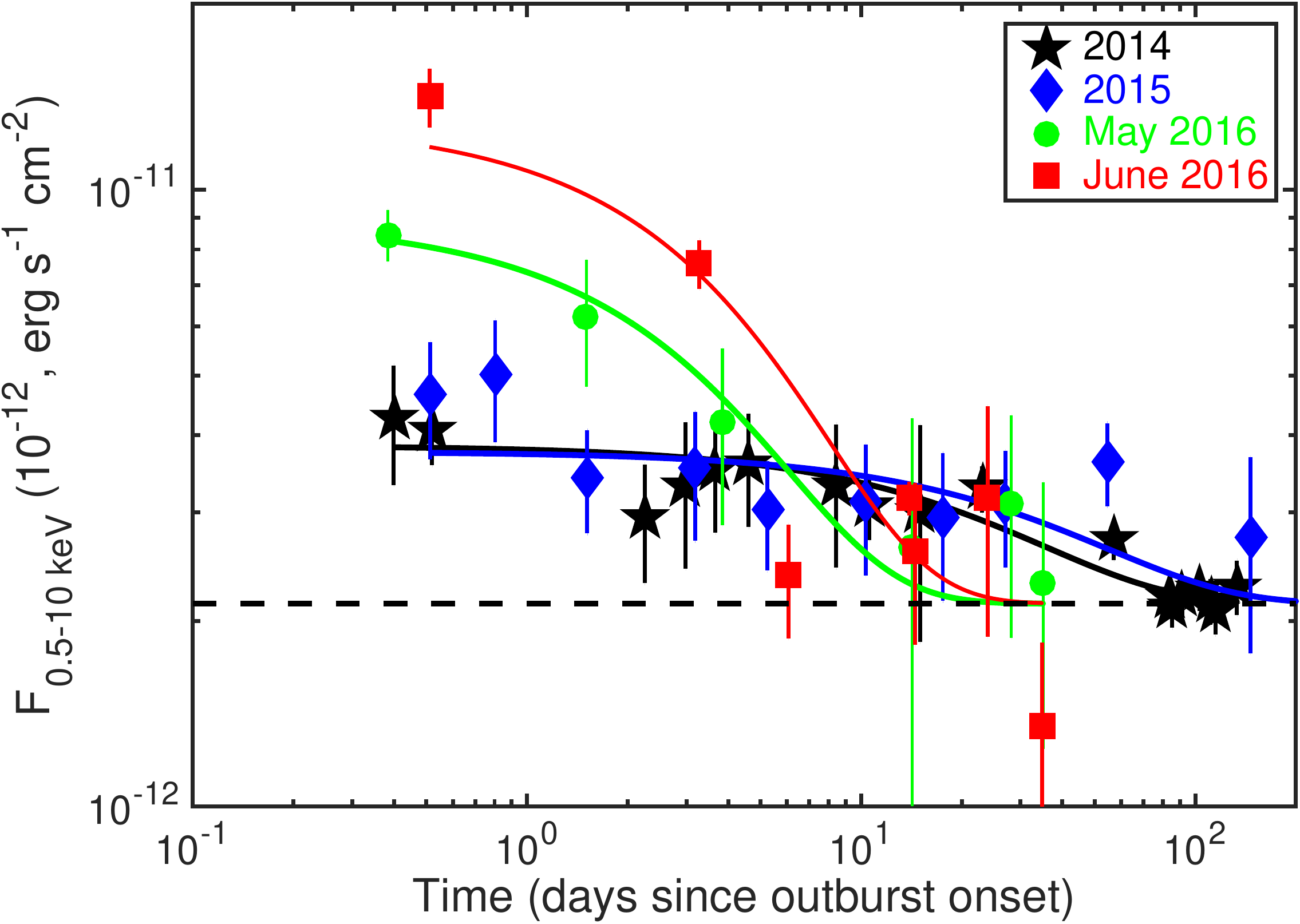}
\caption{Total $0.5-10$~keV flux evolution with time for all four
  outbursts detected from \src. The flux level reached highest at
  outburst onset during the latest outburst of June 2016, during
  which the most number of bursts have been detected from the
  source. Solid lines represent an exponential-decay fit. See text for
  details.}
\label{fluxDecayAll}
\end{center}
\end{figure}

We first concentrate on the 2014 outburst which has the best
observational coverage compared to the rest. The outburst decay is
best fit with an exponential function $F(t)=Ke^{-t/\tau}+F_{\rm q}$,
where $K$ is a normalization factor, while $F_{\rm 
  q}=2.1\times10^{-12}$~erg~s$^{-1}$~cm$^{-2}$ is the 
quiescent flux level derived with the \xmm\ observations
(Figure~\ref{fluxDecayAll}). This fit results in a characteristic
decay time-scale $\tau_{\rm 14}=29\pm4$~days
(Table~\ref{outDecProp}). Integrating over 200 days, we find a total 
energy in the outburst, corrected for the quiescent flux level,
$E_{\rm 14}=(4.1\pm0.7)\times10^{40}$~erg. We find a flux at outburst 
onset $F_{\rm on-14}=(4.3\pm0.7)\times10^{-12}$~erg~s$^{-1}$~cm$^{-2}$, and a
ratio to the quiescent flux level $R_{\rm 14}\approx2.0$. Following
the same recipe for the 2015 outburst, we find a characteristic decay
time-scale $\tau_{\rm 15}=43^{+12}_{-8}$~days, and a total energy in
the outburst, corrected for the quiescent flux level, $E_{\rm
  15}=(6.1\pm1.1)\times10^{40}$~erg. The flux at outburst onset is
$F_{\rm on-15}=(4.7\pm0.08)\times10^{-12}$~erg~s$^{-1}$~cm$^{-2}$,
and its ratio to the quiescent flux level $R_{\rm 15}=2.2$.

A similar analysis for the May and June 2016 outbursts was difficult
to perform due to the lack of observations $\sim30$~days beyond the
start of each outburst (Figure~\ref{fluxDecayAll}), and the poor
constraints on the fluxes (due to the short XRT exposures) derived few
days after the outburst onset. These fluxes are consistent with
$F_{\rm q}$ and the slightly brighter flux level seen in the 2014 and
2015 outbursts between few days after outburst onset and quiescence
reached $\sim70$~days later. Hence, we cannot derive the long term decay
shape of the lightcurve during the last two outbursts from \src.

\begin{table*}[th!]
\caption{Outburst properties}
\label{outDecProp}
\newcommand\T{\rule{0pt}{2.6ex}}
\newcommand\B{\rule[-1.2ex]{0pt}{0pt}}
\begin{center}{
\resizebox{0.8\textwidth}{!}{
\begin{tabular}{l c c c c c c}
\hline
\hline
Outburst \T\B  & $\tau$ & $K$ & $E_{\rm 10}^a$ & $E_{\rm 200}^b$ & $F_{\rm peak}$ & $E_{\rm burst}^c$\\
 \T\B                & (days)   & ($10^{-12}$) & ($10^{40}$~erg) & ($10^{40}$~erg) & ($10^{-12}$~erg~s$^{-1}$~cm$^{-2}$) & ($10^{38}$~erg~s$^{-1}$) \\
\hline
2014  \T\B & $29\pm4$ & $1.7\pm0.2$ &  $1.2\pm0.3$ & $4.1\pm0.7$ & $4.3\pm0.7$ & $8\pm2$\\
2015  \T\B & $43^{+12}_{-8}$ & $1.6_{-0.3}^{+0.4}$ & $1.2\pm0.2$ & $6.1\pm1.1$& $4.7\pm0.8$ & $83\pm3$ \\
2016 May$^d$  \T\B & $3.7\pm1.0$ & $6.8_{-0.5}^{+0.7}$ & $2.0\pm0.3$ & NA & $8.5\pm0.6$ &  $411\pm3$ \\
2016 June$^d$  \T\B & $4.3\pm1.0$ & $10.8_{-2.5}^{+3.2}$ & $3.6\pm0.4$ & NA & $14\pm1$ & $1020\pm8$\\
\hline
\end{tabular}}}
\begin{list}{}{}
\item[{\bf Notes.}] All energies are derived assuming a distance of
  9~kpc. $^a$ Integrated total energy within 10 days from outburst
  onset. $^b$ Integrated total energy within 200 days from outburst
  onset. $^c$ Total energy in the bursts for the day of the
  outburst onset, i.e., 2014 July 05, 2015 February 22, 2016 May 18,
  2016 June 23 (Lin et al. in prep.). $^d$ Long-term outburst behavior
  during 2016 cannot be explored due to lack of high S/N observations
  beyond few days of outburst onset. See text for details.
\end{list}
\end{center}
\end{table*}

However, an exponential decay fit to the 2016 outbursts results in
short term characteristic time-scales $\tau_{\rm
  May-16}\approx\tau_{\rm  June-16}\approx4$~days, indicating a
quick initial decay which might have been followed by a longer one
similar to what is observed in 2014 and 2015. To enable comparison
between all outbursts, we derive the total energy emitted within 10
days of each outburst. These are reported in
Table~\ref{outDecProp}. The 2016 outburst onset to quiescence flux
ratios are $R_{\rm May16}=4.0$ and $R_{\rm June16}=6.7$.
Table~\ref{outDecProp} also includes the total energy in the bursts
during the first day of each of the outbursts ({\sl Lin et al. in
  prep.}).

Finally, we note that the last observation during the May 2016
outburst was taken 1.5~days prior to the start of the June outburst
(last green dot and first red square in
Figure~\ref{fluxDecayAll}). The total fluxes from the two observations
differ at the $\gtrsim5\sigma$ level. These results are discussed in
Section~\ref{outDec}.

\section{Discussion}
\label{discuss}

\subsection{Broad-band X-ray properties}

%
%
\font\fiverm=cmr5       \font\fivebf=cmbx5      \font\sevenrm=cmr7
\font\sevenbf=cmbx7     \font\eightrm=cmr8      \font\eighti=cmmi8
\font\eightsy=cmsy8     \font\eightbf=cmbx8     \font\eighttt=cmtt8
\font\eightit=cmti8     \font\eightsl=cmsl8     \font\sixrm=cmr6
\font\sixi=cmmi6        \font\sixsy=cmsy6       \font\sixbf=cmbx6

\def\edithere#1{\textcolor{red}{\bf #1}}  
\def\actionitem#1{\textcolor{blue}{\bf #1}}  

\def\rns{R_{\hbox{\sixrm NS}}}
\def\r\Ans{A_{\hbox{\sixrm NS}}}

Using high S/N ratio observations, we have established that the \src\
soft X-ray spectrum, with photon energies $<10$~keV, is well described
with the phenomenological BB+PL or 2BB model. \nustar\ observations,
on the other hand, were crucial in providing the first look at this
magnetar at energies $>10$~keV, revealing a hard X-ray tail extending
up to $79~$keV. We note that this \nustar\ observation was taken 5
days after the 2015 outburst. The simultaneous \swift+\nustar\ fit
revealed a $0.5-10$~keV flux $\sim40\%$ larger than the quiescent
flux, which we assume it to be at the 2014 \xmm\ level of
$2.2\times10^{-12}$~erg~s$^{-1}$~cm$^{-2}$. The spectra below 10~keV
did not show significant spectral variability during any of the
outbursts (Section~\ref{outDec}), except for the relative
brightness. Accordingly, one can conjecture that \src\ has a similar
high-energy tail during quiescence, though proof of such requires
further dedicated monitoring of the source with NuSTAR or INTEGRAL.

The presence of hard-X-ray tails, such as exhibited by \src, is
clearly seen in about a third of all known magnetars \citep[e.g.,
][]{kuiper06ApJ,hartog08AA:0142,enoto10ApJ,esposito07AA:1806},
but may indeed be universal to them. Spectral details differ across the
population. For instance, the hard X-ray tail photon index we measure,
$\Gamma_{\rm H}\approx0.9$, is quite similar to some measured for AXPs
\citep[e.g., ][]{an13ApJ:1841, vogel14ApJ:2259,tendulkar15ApJ:0142,
  denhartog08AA:1rxs1708}, but somewhat harder than other sources
\citep[e.g., ][]{esposito07AA:1806,yang16ApJ:1048}. Moreover, the flux
in the hard PL tail is 2 times larger than the flux in the soft
components. This flux ratio varies by about 2 orders of magnitude 
among the magnetar population \citep{enoto10ApJ}.

\citet[][see also \citealt{marsden01ApJ,enoto10ApJ}]{kaspi10ApJ}
searched for correlations between the observed X-ray parameters and
the intrinsic parameters for magnetars. They found an anti-correlation
between the index differential $\Gamma_{\rm S}-\Gamma_{\rm H}$
and the strength of the magnetic field $B$. For \src, with its
spin-down field strength of $B=2.2\times10^{14}$~G
\citep{israel16mnras:1935}, the determination here of $\Gamma_{\rm
  S}-\Gamma_{\rm H}\approx1.0-2.0$ nicely fits the \cite{kaspi10ApJ}
correlation. Moreover, \citet{enoto10ApJ} noted a strong correlation
between the hardness ratio, defined as $F_{\rm H}/F_{\rm S}$ for the
hard and soft energy bands, respectively, and the characteristic age
$\tau$. Following the same definition for the energy bands as in
\citet{enoto10ApJ}, we find $F_{\rm H}/F_{\rm S}\approx1.8$ which
falls very close to this correlation line given the \src\ spin-down
age $\tau=3.6$~Kyr \citep{israel16mnras:1935}. Since the electric
field for a neutron star $E$ along its last open field line is
nominally inversely proportional to the characteristic spin down age
$E=\Omega RB\propto \tau^{-1/2}$, \citet{enoto10ApJ} argued that a
younger magnetar will be able to sustain a larger current,
accelerating more particles into the magnetosphere and causing a
stronger hard X-ray emission in the tail. This scenario is predicated
on the conventional picture of powerful young rotation-powered pulsars
like the Crab.

The most discussed model for generating a hard X-ray tail in magnetar
spectra is resonant Compton up-scattering of soft thermal photons by
highly relativistic electrons with Lorentz factors $\sim10-10^4$ in
the stellar magnetosphere \citep[e.g.,][]{baring07,fernandez07ApJ,
  beloborodov13ApJ}. The emission locale is believed to be at
distances $\sim10-100 \rns$ where $\rns=10~$km is the neutron star
radius. There the intense soft X-ray photon field seeds the inverse
Compton mechanism, and the collisions are prolific because of
scattering resonances at the cyclotron frequency and its harmonics in
the rest frame of an electron. Magnetar conditions guarantee that
electrons accelerated by voltages in the inner magnetosphere will cool
rapidly down to Lorentz factors $\gamma\sim10-10^2$ \citep{bwg11ApJ}
due to the resonant scatterings. Along each field line, the
up-scattered spectra are extremely flat, with indices $\Gamma_h\sim
-0.5$ -- $0.0$ \citep[][see also Wadiasingh, et al., in
prep.]{baring07}, though the convolution of contributions from
extended regions is necessarily steeper, and more commensurate with
the observed hard tail spectra \citep{beloborodov13ApJ}. While the
inverse Compton emission can also extend out to gamma-ray energies,
the prolific action of attenuation mechanisms such as magnetic pair
creation $\gamma \to e^+e^-$ and photon splitting
$\gamma\to\gamma\gamma$ \citep{baring01ApJ} limits emergent signals to
energies below a few MeV in magnetars \citep{story14ApJ}, and probably
even below 500 keV.

\citet[][see also \citealt{chen16:mag}]{beloborodov13ApJ} developed a
coronal outflow model based on the above picture, using the twisted
magnetosphere scenario \citep{thompson02ApJ:magnetars,
  beloborodov09ApJ}. Twists in closed magnetic field loops (dubbed
$J$-bundles) extending high into the magnetosphere can accelerate
particles to high Lorentz factors, which will decelerate and lose
energy via resonant Compton up-scattering. If pairs are created in
profusion, they then annihilate at the top of a field loop. Another
one of the $J$-bundle model predictions is a hot spot on the surface
formed when return currents hit the surface at the footprint of the
twisted magnetic field lines. The physics in this model is mostly
governed by the field lines twist amplitude $\psi$ \citep{
  thompson02ApJ:magnetars}, the voltage $\Phi_j$ in the bundle, and
its half-opening angle to the magnetic axis $\theta_{\rm j}$
\citep{beloborodov13ApJ,hascoet14ApJ}.

The temperatures expected for the hotspots are of the order of
$\sim1$~keV while areas depend on the geometry of the bundle and
the angle $\theta_j$. For a dipole geometry, $A_{\rm
  j}\sim(1/4)\theta_{\rm j}^2A_{\rm ns}\approx0.02(\theta_{\rm
  j}/0.3)^2A_{\rm ns}$, where $A_{\rm ns}=4\pi R_{\odot}^2$ is the NS
surface area \citep{hascoet14ApJ}. Assuming that the hot BB in our
model discussed in the last paragraph of Section~\ref{SecOut2014}
represents the footprints of the $J$-bundle, for which we find a
temperature $kT=0.8$~keV, we estimate its surface area
$A\approx0.6$~km$^2$. Assuming that $A\approx A_{\rm j}$, we estimate
$\theta_{\rm j}\approx0.05$. 

The above calculation assumes that the $J$-bundle is axisymmetric
extending all around the NS. The hot-spot, hence, is a ring around the
polar cap rim. The smaller area that we derive may suggest that the
$J$-bundle is not axisymmetric and extends only around part of the NS,
implying that the twist could have been imparted onto local magnetic
field lines.



The total power dissipated by the $J$-bundle in the twisted
magnetosphere model can be expressed as $L_{\rm
  j}\approx2\times10^{35}\psi\Phi_{\rm 10}\mu_{\rm 32}R_{\rm
  10}\theta_{\rm j,0.3}^4$~erg~s$^{-1}$ (equation 3,
\citealt{hascoet14ApJ}), where $\Phi_{\rm 10}$ is the voltage in 
units of $10^{10}$~V, $\mu_{\rm 32}$ is the magnetic moment in units
of $10^{32}$~G~cm$^{3}$, $R_{\rm 10}$ is the NS radius in units of
10~km, and $\theta_{\rm j,0.3}=\theta_{\rm j}/0.3$. Given the magnetic
moment of \src, for choices of $\phi_{\rm 10}=1$, $\psi=1$, $R_{\rm
  10}=1$, and $\theta_{\rm j,0.3}\approx0.2$, we estimate $L_{\rm
  j}=7\times10^{32}$~erg~s$^{-1}$. This luminosity is a factor of
$\sim30$ smaller than the hard tail PL luminosity, $L_{\rm
  PL}=2.0\times10^{34}$~erg~s$^{-1}$ we derive with the \nustar\ data, 
after normalizing it to the 2014 \xmm\ flux level\footnote{The
  \nustar\ observation was taken 5 days after the outburst when the
  simultaneous XRT observation showed an increase in the PL flux by a
  factor of 2 above the quiescent \xmm\ level of 2014. We normalized the
  hard PL luminosity from table~\ref{specParamNus} by the same
  factor. See also footnote~5.}. This might imply a larger voltage
across the twisted field lines than the choice of $\phi_{10}=1$, which
corresponds to only $\sim3\times 10^{-6}$ times the open field line
pole-to-equator voltage $2\pi \rns^2 B_p/(Pc)\approx 2.8\times
10^{16}$~V for \src. Another possibility is that the hard PL tail
could be much fainter during quiescence, which might indicate a
different decay trend for the high-energy tail compared to the
0.5-10~keV spectrum. A deep \xmm+\nustar\ observation of \src\ during
quiescence would help reveal the exact shape and power of the hard PL
tail, inform on how activation relates to heat transfer to and from
the stellar surface layers, and help refine the twisted magnetosphere
model.

\subsection{Outbursts}
\label{outDec}

Since its discovery in June 2014, \src\ has shown four major 
bursting episodes, which culminated with the strongest one to date in
June 2016. Similar to most other magnetars, \src\ bursting activity
was accompanied by a persistent emission outburst, showing an
increase in the flux level at, or shortly after, the onset of the
bursting activity that decayed quasi-exponentially back to quiescence
\citep[e.g., ][]{woods04ApJ:1E2259,scholz12ApJ:1822,gogus10ApJ:1833,
  kargaltsev12apj:1834,zelati15MNRAS:1745,younes15ApJ:1806,rea11:outburst}.

The rise time of magnetar outbursts is a challenging observational
property to identify and quantify due to the randomness of the
process. Magnetars are usually observed by pointed X-ray telescopes
after they have entered a bursting episode. Hence, it is unclear
whether magnetars show any persistent flux enhancement prior to the
bursting activity, or whether the two happen (quasi-) simultaneously.
CXOU~J164710.2$-$455216 is the closest we have come to answering the
above question. While being monitored with X-ray instruments,
CXOU~J164710.2$-$455216 was observed with \xmm\ 5~days prior to
bursting activity \citep{israel07ApJ:1647}. The flux of this
observation was consistent with quiescence while the following
observation, which took place less than a day after the bursts, showed
an increase by a factor of $\sim300$. Similarly, \src\ was observed
$\sim$1.5~days prior to its strongest bursting activity in 2016
June, while being monitored for its 2016 May activation. The latter
observation showed a flux level close to quiescence, and was $5\sigma$
away from the flux measured at the start of the June 2016 outburst 
(Section~\ref{outComapre}). This implies that any instability invoked
to explain the outbursts in magnetars has to develop on very short
time-scales ($\lesssim2$~days, e.g., \citealt{li16ApJ:HW}).

The $0.5-10$~keV persistent flux level of \src\ at or shortly after the
onset of the bursting activity varied in concordance with the bursting
level from the source \citep[see also, e.g., 1E 1547.0-5408, ][]{
  ng11ApJ:1547}. The source flux reached its highest level at the start of the 2016
June outburst, a factor of 7 that of the quiescent level
(Figure~\ref{fluxDecayAll}). At the same time, the flux of the PL
or the hot BB components (Figures \ref{specEvolPL} and
\ref{specEvol2BB}), increased by a factor of $\sim25$ compared to quiescence. 
The cold BB on the other hand, with a temperature of $kT=0.48$~keV and radius
$R=1.8$~km, remained more or less constant throughout all four
outbursts. Such a cold BB could be the result of internal heating of a
large fraction of the magnetar surface \citep{thompson96ApJ:magnetar,
  beloborodov16:heat}.

The 2014 and 2015 flux decays followed a simple exponential trend with
time-scales of  $\sim30-40$~days. The brighter 2016 outbursts,
however, exhibited a quick decay trend on time-scales of
$\sim4$~days. Such fast initial drop in flux is seen at the outburst
onset of a number of magnetars (e.g., SGR~J1627$-$41,
\citealt{an12ApJ:1627}; Swift~J1834.9$-$0846, \citealt{ 
  kargaltsev12apj:1834}, Swift~J1822.3$-$1606, \citealt{
  scholz12ApJ:1822}). 

Similar amount of energy was emitted in the 2014 and 2015 outbursts
(within $2\sigma$), $E\sim5\times10^{40}$~erg~s$^{-1}$. We were only
able to quantify the total energy emitted during the first 10 days of
the May and June 2016 outbursts, $E=2\times10^{40}$~erg~s$^{-1}$ and
$E=3.6\times10^{40}$~erg~s$^{-1}$, respectively. The energetics in
these outbursts are at the lower end compared to the bulk of magnetar
outbursts \citep{rea11:outburst}. We note that the energy in the
bursts for the 4 outbursts varied by more than 2 orders of magnitude
(Table~\ref{outDecProp}, {\it Lin et al. 2017 in prep.}); a much
larger increase than the energy emitted  in the outbursts. For
instance, the 2014 and 2015 ratio of total energy in the outbursts to
total energy in bursts decreased from 50 to 8.

\begin{figure}[!h]
\begin{center}
\includegraphics[angle=0,width=0.49\textwidth]{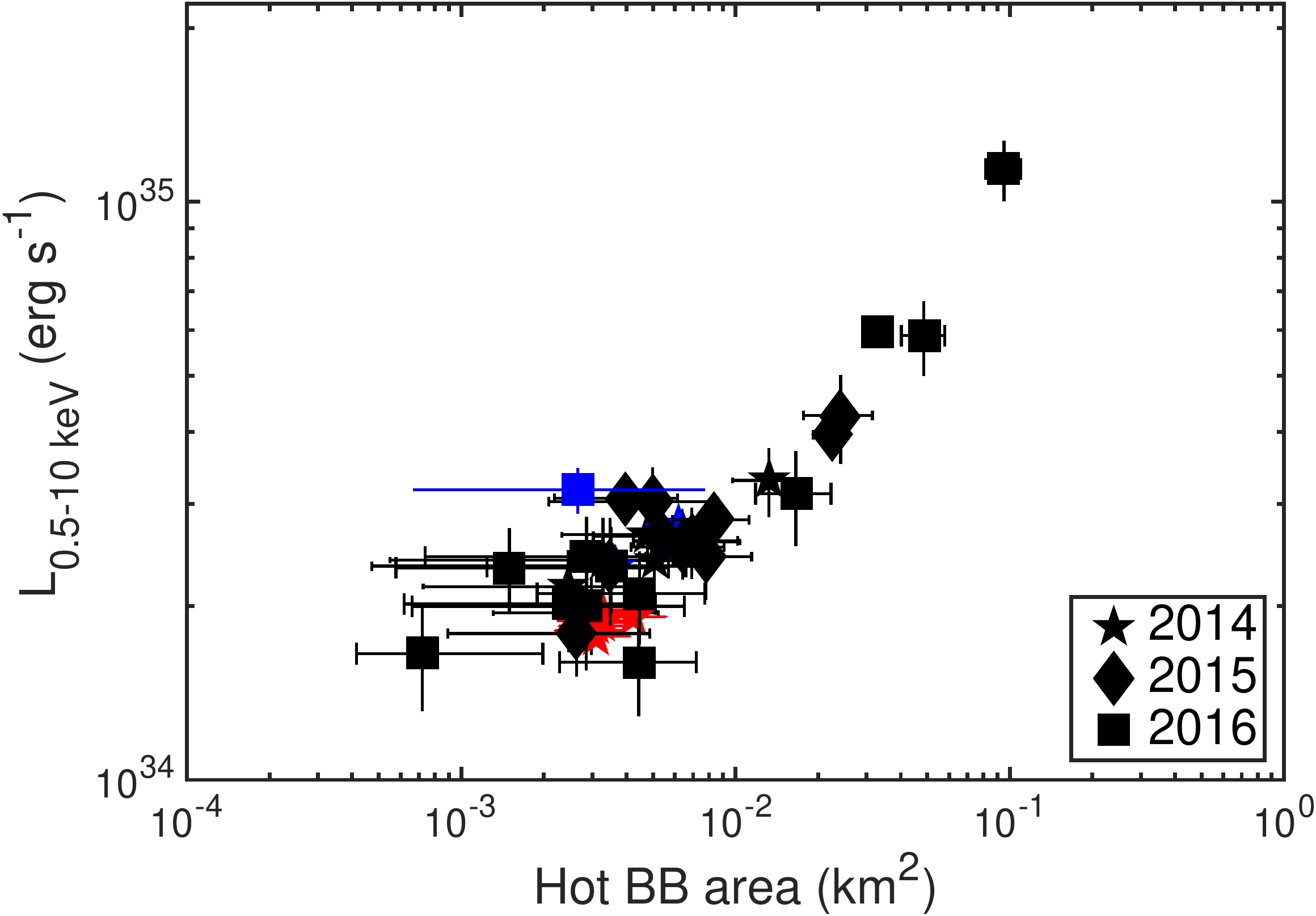}
\caption{Total $0.5-10$~keV flux versus hot BB area from all outbursts
  of \src. See also Figure~\ref{specEvol2BB}.}
\label{FvsA}
\end{center}
\end{figure}

Two models have been discussed in the context of magnetar
outbursts. The first invoked an instability (external or internal)
that rapidly (within few days) deposits energy, of the order of
$10^{40}-10^{42}$~erg~s$^{-1}$, at the crust level of the neutron star
\citep[e.g.,][]{lyubarski02ApJ:magcool,pons12ApJ:mag,brown09ApJ:heat}.
The depth at which the heat is deposited governs the outburst decay
time-scale, which can range from weeks to months, as the crust cools
back to its pre-outburst level. This timescale may also reflect the
magnetic colatitude of the energy dissipation locale, as heat
conductivity across strong fields is suppressed, so that vertical
transport of energy is easier in polar activation zones. This picture
fits the observed properties of the 2014 and 2015 outbursts of
\src. It is, however, difficult to reconcile the initial quick decay
of a few days observed in the 2016 outbursts, and a number of other
magnetars, with this model.

In the second theoretical picture, magnetar outbursts are believed to
be triggered when stresses on the crust build up to a critical level
due to Hall wave propagation caused by magnetic field evolution inside
the NS \citep[e.g., ][]{thompson96ApJ:magnetar,pons12ApJ:mag,
  li16ApJ:HW}. These stresses twist a bundle of external magnetic field
lines anchored to the surface, accelerating particles off the surface
of the star, while returning currents deposit heat at the footprints
of these lines (hot spot, \citealt{thompson02ApJ:magnetars,
  beloborodov09ApJ}). This instability develops on days to weeks
time-scale \citep{li16ApJ:HW}, with decay time-scales ranging from
weeks to years and primarily depending on the strength of the twist
imparted onto the B-field bundle. These properties match the outburst
properties that we observe for \src. Another prediction of this model
is a shrinking hot spot at the surface, which we do observe when we
fit the $0.5-10$~keV spectra with the 2BB model\footnote{We do not
  attempt to quantitatively compare the Flux versus Area relation we
  observe here to the prediction of \citet{beloborodov09ApJ} due to
  uncertainties in the parameter estimates of the 2BB model as
  discussed in Section~\ref{SecOut2014}.} (Figure~\ref{FvsA}).
However, similar to the crust heating model, it is not trivial to
explain the initial quick decay observed in the 2016 outbursts with the
twisted magnetosphere model.

\subsection{Radio comparison to other magnetars}

The upper limits on the radio counterpart that we have obtained are 
the deepest radio limits for \src\ thus far (i.e.,
\citealt{surnis2016ApJ:1935}). In fact, our Arecibo observations
represent the deepest radio observations that were carried out quickly
after the X-ray outburst of a magnetar,
\citep[e.g.,][]{crawford2007,lazarus2012}. Currently it is not clear
what is the best epoch to search for magnetar radio emission. The
sample of magnetars with radio detections is small, and although some
were detected close in time to their X-ray activation, there does not
seem to be a clear correlation between magnetar X-ray and radio
activity. We note in this context that the \src\ spindown luminosity
of $1.7\times10^{34}$~erg/s and X-ray luminosity in quiescence of
$2.1\times10^{34}$~erg/s ($0.5-10$~keV) put \src\ in the area of
magnetars that are not expected to display radio emission in the
fundamental plane of \citet{rea12ApJ:radiomag}. The latter needs to be
tested further with deep radio searches as presented in this paper,
both when magnetars are X-ray active, as well as when they are in
their quiescent state.

\section{Conclusion}
\label{sum}

In the following we summarize the main findings of our analyses of the
broad-band X-ray and radio data of the magnetar \src\ taken in the
aftermath of its 2014, 2015, and 2016 outbursts:

\begin{itemize}[noitemsep]
\item \chandra\ data did not reveal any small-scale extended emission
  around \src.
\item No pulsations are detected from \src\ in the days following the
  2015 and 2016 outbursts. We derive an upper-limit of 25\% and 8\% in
  the energy range $3-50$~keV during 2015, and $1-8$~keV during 2016,
  with \nustar\ and \chandra\ respectively.
\item No radio pulsations are detected with Arecibo from \src\
  following the 2014 and 2016 outbursts. We set the deepest limits on
  the radio emission from a magnetar, with a maximum flux density
  limit of $14$\,$\mu$Jy for the 4.6~GHz observations and $7$\,$\mu$Jy
  for the 1.4 GHz observations.
\item The soft X-ray spectrum $<10$~keV is well described with a BB+PL
  or 2BB model during all three outbursts.
\item \nustar\ observations 5 days after the 2015 outburst onset
  revealed a hard X-ray tail, $\Gamma=0.9$, extending up to $79~$keV,
  with flux larger than the one detected $<10$~keV.
\item Following the outbursts, the $0.5-10$~keV flux from \src\
  increased in concordance to its bursting activity. At the onset of
  the 2016 June bursting episode, the strongest one to date, the
  $0.5-10$~keV reached maximum, increasing by a factor of $\sim7$
 above its quiescent level.
\item The $0.5-10$~keV flux increase during the outbursts is due to
  the PL or hot BB component, which increased by a maximum factor of
  $25$ compared to quiescence. The cold BB component, $kT=0.47$~keV,
  remained more or less constant. 
\item The 2014 and 2015 outbursts decayed quasi-exponentially with
  time-scales of $\sim40$~days. The stronger May and June 2016
  outbursts showed a quick short-term decay with time-scales of
  $\sim4$~days; their long-term decay trends were not possible to
  derive.
\item The last \swift/XRT observation of the May 2016 outburst, taken
  $1.5$ days prior to the onset of the 2016 June outburst, showed a
  flux level close to quiescence, and was dimmer at the $5\sigma$
  level compared to the flux measured at the start of the June 2016
  outburst.
\item The total energy emitted by the bursts increased by two orders
  of magnitude between the 2014 and the 2016 June outbursts
  (Table~\ref{outDecProp}, {\it Lin et al. 2017 in prep.}). This is a
  much larger increase compared to the energy emitted by the star
  through the increase of its X-ray persistent emission.
\end{itemize}

\section*{Acknowledgments}

We thank \nustar\ PI Fiona Harrison and Belinda Wilkes for granting
\nustar\ and \chandra\ DDT observations of \src\ during the 2015 and
june 2016 outbursts, respectively. We also thank the \swift\ team for
performing the monitoring of the source during all of its
outbursts. G.Y. and C.K. acknowledge support  by  NASA  through  grant
NNH07ZDA001-GLAST. A.J. and J.W.T.H. acknowledge funding from the
European Research Council under the European Union's Seventh Framework
Programme (FP7/2007- 2013) ERC grant agreement nr. 337062
(DRAGNET). The Arecibo Observatory is operated by SRI International
under a cooperative agreement with the National Science Foundation
(AST-1100968), and in alliance with Ana G. M\'{e}ndez-Universidad
Metropolitana, and the Universities Space Research Association. We
would like to thank Arecibo observatory scheduler Hector Hernandez for
the support during our observations.


\begin{thebibliography}{89}
\expandafter\ifx\csname natexlab\endcsname\relax\def\natexlab#1{#1}\fi

\bibitem[{{An} {et~al.}(2013){An}, {Hasco{\"e}t}, {Kaspi}, {Beloborodov},
  {Dufour}, {Gotthelf}, {Archibald}, {Bachetti}, {Boggs}, {Christensen},
  {Craig}, {Greffenstette}, {Hailey}, {Harrison}, {Kitaguchi}, {Kouveliotou},
  {Madsen}, {Markwardt}, {Stern}, {Vogel}, \& {Zhang}}]{an13ApJ:1841}
{An}, H., {Hasco{\"e}t}, R., {Kaspi}, V.~M., {et~al.} 2013, \apj, 779, 163

\bibitem[{{An} {et~al.}(2012){An}, {Kaspi}, {Tomsick}, {Cumming}, {Bodaghee},
  {Gotthelf}, \& {Rahoui}}]{an12ApJ:1627}
{An}, H., {Kaspi}, V.~M., {Tomsick}, J.~A., {et~al.} 2012, \apj, 757, 68

\bibitem[{{Antonopoulou} {et~al.}(2015){Antonopoulou}, {Weltevrede},
  {Espinoza}, {Watts}, {Johnston}, {Shannon}, \&
  {Kerr}}]{Antonopoulou2015MNRAS:1119}
{Antonopoulou}, D., {Weltevrede}, P., {Espinoza}, C.~M., {et~al.} 2015, \mnras,
  447, 3924

\bibitem[{{Archibald} {et~al.}(2016){Archibald}, {Kaspi}, {Tendulkar}, \&
  {Scholz}}]{archibald16:j1119}
{Archibald}, R.~F., {Kaspi}, V.~M., {Tendulkar}, S.~P., \& {Scholz}, P. 2016,
  ArXiv e-prints

\bibitem[{{Arnaud}(1996)}]{arnaud96conf}
{Arnaud}, K.~A. 1996, in ASP Conf. Ser. 101: Astronomical Data Analysis
  Software and Systems V, 17

\bibitem[{{Baring} \& {Harding}(1998)}]{baring98ApJ}
{Baring}, M.~G. \& {Harding}, A.~K. 1998, \apjl, 507, L55

\bibitem[{{Baring} \& {Harding}(2001)}]{baring01ApJ}
{Baring}, M.~G. \& {Harding}, A.~K. 2001, \apj, 547, 929

\bibitem[{{Baring} \& {Harding}(2007)}]{baring07}
{Baring}, M.~G. \& {Harding}, A.~K. 2007, \apss, 308, 109

\bibitem[{{Baring} {et~al.}(2011){Baring}, {Wadiasingh}, \&
  {Gonthier}}]{bwg11ApJ}
{Baring}, M.~G., {Wadiasingh}, Z., \& {Gonthier}, P.~L. 2011, \apj, 733, 61

\bibitem[{{Beloborodov}(2009)}]{beloborodov09ApJ}
{Beloborodov}, A.~M. 2009, \apj, 703, 1044

\bibitem[{{Beloborodov}(2013)}]{beloborodov13ApJ}
{Beloborodov}, A.~M. 2013, \apj, 762, 13

\bibitem[{{Beloborodov} \& {Li}(2016)}]{beloborodov16:heat}
{Beloborodov}, A.~M. \& {Li}, X. 2016, ArXiv e-prints

\bibitem[{{Bhattacharya}(1998)}]{Bhat:1998}
{Bhattacharya}, D. 1998, in NATO Advanced Science Institutes (ASI) Series C,
  Vol. 515, NATO Advanced Science Institutes (ASI) Series C, ed. R.~{Buccheri},
  J.~{van Paradijs}, \& A.~{Alpar}, 103

\bibitem[{{Brown} \& {Cumming}(2009)}]{brown09ApJ:heat}
{Brown}, E.~F. \& {Cumming}, A. 2009, \apj, 698, 1020

\bibitem[{{Buccheri} {et~al.}(1983){Buccheri}, {Bennett}, {Bignami}, {Bloemen},
  {Boriakoff}, {Caraveo}, {Hermsen}, {Kanbach}, {Manchester}, {Masnou},
  {Mayer-Hasselwander}, {Ozel}, {Paul}, {Sacco}, {Scarsi}, \&
  {Strong}}]{buccheri83AApulse}
{Buccheri}, R., {Bennett}, K., {Bignami}, G.~F., {et~al.} 1983, \aap, 128, 245

\bibitem[{{Burrows} {et~al.}(2005){Burrows}, {Hill}, {Nousek}, {Kennea},
  {Wells}, {Osborne}, {Abbey}, {Beardmore}, {Mukerjee}, {Short}, {Chincarini},
  {Campana}, {Citterio}, {Moretti}, {Pagani}, {Tagliaferri}, {Giommi},
  {Capalbi}, {Tamburelli}, {Angelini}, {Cusumano}, {Br{\"a}uninger}, {Burkert},
  \& {Hartner}}]{burrows05SSRv:xrt}
{Burrows}, D.~N., {Hill}, J.~E., {Nousek}, J.~A., {et~al.} 2005, \ssr, 120, 165

\bibitem[{{Camilo} {et~al.}(2007){Camilo}, {Ransom}, {Halpern}, \&
  {Reynolds}}]{camilo07ApJ:1550}
{Camilo}, F., {Ransom}, S.~M., {Halpern}, J.~P., \& {Reynolds}, J. 2007, \apjl,
  666, L93

\bibitem[{{Camilo} {et~al.}(2006){Camilo}, {Ransom}, {Halpern}, {Reynolds},
  {Helfand}, {Zimmerman}, \& {Sarkissian}}]{camilo06Natur:1810}
{Camilo}, F., {Ransom}, S.~M., {Halpern}, J.~P., {et~al.} 2006, \nat, 442, 892

\bibitem[{{Chen} \& {Beloborodov}(2016)}]{chen16:mag}
{Chen}, A.~Y. \& {Beloborodov}, A.~M. 2016, ArXiv e-prints

\bibitem[{{Coti Zelati} {et~al.}(2015){Coti Zelati}, {Rea}, {Papitto},
  {Vigan{\`o}}, {Pons}, {Turolla}, {Esposito}, {Haggard}, {Baganoff}, {Ponti},
  {Israel}, {Campana}, {Torres}, {Tiengo}, {Mereghetti}, {Perna}, {Zane},
  {Mignani}, {Possenti}, \& {Stella}}]{zelati15MNRAS:1745}
{Coti Zelati}, F., {Rea}, N., {Papitto}, A., {et~al.} 2015, \mnras, 449, 2685

\bibitem[{{Crawford} {et~al.}(2007){Crawford}, {Hessels}, \&
  {Kaspi}}]{crawford2007}
{Crawford}, F., {Hessels}, J.~W.~T., \& {Kaspi}, V.~M. 2007, \apj, 662, 1183

\bibitem[{{den Hartog} {et~al.}(2008{\natexlab{a}}){den Hartog}, {Kuiper}, \&
  {Hermsen}}]{denhartog08AA:1rxs1708}
{den Hartog}, P.~R., {Kuiper}, L., \& {Hermsen}, W. 2008{\natexlab{a}}, \aap,
  489, 263

\bibitem[{{den Hartog} {et~al.}(2008{\natexlab{b}}){den Hartog}, {Kuiper},
  {Hermsen}, {Kaspi}, {Dib}, {Kn{\"o}dlseder}, \& {Gavriil}}]{hartog08AA:0142}
{den Hartog}, P.~R., {Kuiper}, L., {Hermsen}, W., {et~al.} 2008{\natexlab{b}},
  \aap, 489, 245

\bibitem[{{Dewey} {et~al.}(1985){Dewey}, {Taylor}, {Weisberg}, \&
  {Stokes}}]{DTW:1985}
{Dewey}, R.~J., {Taylor}, J.~H., {Weisberg}, J.~M., \& {Stokes}, G.~H. 1985,
  \apjl, 294, L25

\bibitem[{{Enoto} {et~al.}(2010){Enoto}, {Nakazawa}, {Makishima}, {Rea},
  {Hurley}, \& {Shibata}}]{enoto10ApJ}
{Enoto}, T., {Nakazawa}, K., {Makishima}, K., {et~al.} 2010, \apjl, 722, L162

\bibitem[{{Esposito} {et~al.}(2011){Esposito}, {Israel}, {Turolla}, {Mattana},
  {Tiengo}, {Possenti}, {Zane}, {Rea}, {Burgay}, {G{\"o}tz}, {Mereghetti},
  {Stella}, {Wieringa}, {Sarkissian}, {Enoto}, {Romano}, {Sakamoto},
  {Nakagawa}, {Makishima}, {Nakazawa}, {Nishioka}, \& {Fran{\c
  c}ois-Martin}}]{esposito11MNRAS:1833}
{Esposito}, P., {Israel}, G.~L., {Turolla}, R., {et~al.} 2011, \mnras, 416, 205

\bibitem[{{Esposito} {et~al.}(2007){Esposito}, {Mereghetti}, {Tiengo}, {Zane},
  {Turolla}, {G{\"o}tz}, {Rea}, {Kawai}, {Ueno}, {Israel}, {Stella}, \&
  {Feroci}}]{esposito07AA:1806}
{Esposito}, P., {Mereghetti}, S., {Tiengo}, A., {et~al.} 2007, \aap, 476, 321

\bibitem[{{Fern{\'a}ndez} \& {Thompson}(2007)}]{fernandez07ApJ}
{Fern{\'a}ndez}, R. \& {Thompson}, C. 2007, \apj, 660, 615

\bibitem[{{Gavriil} {et~al.}(2008){Gavriil}, {Gonzalez}, {Gotthelf}, {Kaspi},
  {Livingstone}, \& {Woods}}]{gavriil08Sci:psr1846}
{Gavriil}, F.~P., {Gonzalez}, M.~E., {Gotthelf}, E.~V., {et~al.} 2008, Science,
  319, 1802

\bibitem[{{G{\"o}{\u g}{\"u}{\c s}} {et~al.}(2010){G{\"o}{\u g}{\"u}{\c s}},
  {Cusumano}, {Levan}, {Kouveliotou}, {Sakamoto}, {Barthelmy}, {Campana},
  {Kaneko}, {Stappers}, {de Ugarte Postigo}, {Strohmayer}, {Palmer}, {Gelbord},
  {Burrows}, {van der Horst}, {Mu{\~n}oz-Darias}, {Gehrels}, {Hessels},
  {Kamble}, {Wachter}, {Wiersema}, {Wijers}, \& {Woods}}]{gogus10ApJ:1833}
{G{\"o}{\u g}{\"u}{\c s}}, E., {Cusumano}, G., {Levan}, A.~J., {et~al.} 2010,
  \apj, 718, 331

\bibitem[{{G{\"o}{\u g}{\"u}{\c s}} {et~al.}(2016){G{\"o}{\u g}{\"u}{\c s}},
  {Lin}, {Kaneko}, {Kouveliotou}, {Watts}, {Chakraborty}, {Alpar},
  {Huppenkothen}, {Roberts}, {Younes}, \& {van der Horst}}]{gogus16:j1119}
{G{\"o}{\u g}{\"u}{\c s}}, E., {Lin}, L., {Kaneko}, Y., {et~al.} 2016, \apjl,
  829, L25

\bibitem[{{G{\"o}{\v g}{\"u}{\c s}} {et~al.}(2002){G{\"o}{\v g}{\"u}{\c s}},
  {Kouveliotou}, {Woods}, {Finger}, \& {van der Klis}}]{gogus02ApJ:1806}
{G{\"o}{\v g}{\"u}{\c s}}, E., {Kouveliotou}, C., {Woods}, P.~M., {Finger},
  M.~H., \& {van der Klis}, M. 2002, \apj, 577, 929

\bibitem[{{Granot} {et~al.}(2016){Granot}, {Gill}, {Younes}, {Gelfand},
  {Harding}, {Kouveliotou}, \& {Baring}}]{granot16:1834}
{Granot}, J., {Gill}, R., {Younes}, G., {et~al.} 2016, ArXiv e-prints

\bibitem[{{Harrison} {et~al.}(2013){Harrison}, {Craig}, {Christensen},
  {Hailey}, {Zhang}, {Boggs}, {Stern}, {Cook}, {Forster}, {Giommi},
  {Grefenstette}, {Kim}, {Kitaguchi}, {Koglin}, {Madsen}, {Mao}, {Miyasaka},
  {Mori}, {Perri}, {Pivovaroff}, {Puccetti}, {Rana}, {Westergaard}, {Willis},
  {Zoglauer}, {An}, {Bachetti}, {Barri{\`e}re}, {Bellm}, {Bhalerao},
  {Brejnholt}, {Fuerst}, {Liebe}, {Markwardt}, {Nynka}, {Vogel}, {Walton},
  {Wik}, {Alexander}, {Cominsky}, {Hornschemeier}, {Hornstrup}, {Kaspi},
  {Madejski}, {Matt}, {Molendi}, {Smith}, {Tomsick}, {Ajello}, {Ballantyne},
  {Balokovi{\'c}}, {Barret}, {Bauer}, {Blandford}, {Brandt}, {Brenneman},
  {Chiang}, {Chakrabarty}, {Chenevez}, {Comastri}, {Dufour}, {Elvis}, {Fabian},
  {Farrah}, {Fryer}, {Gotthelf}, {Grindlay}, {Helfand}, {Krivonos}, {Meier},
  {Miller}, {Natalucci}, {Ogle}, {Ofek}, {Ptak}, {Reynolds}, {Rigby},
  {Tagliaferri}, {Thorsett}, {Treister}, \& {Urry}}]{harrison13ApJ:NuSTAR}
{Harrison}, F.~A., {Craig}, W.~W., {Christensen}, F.~E., {et~al.} 2013, \apj,
  770, 103

\bibitem[{{Hasco{\"e}t} {et~al.}(2014){Hasco{\"e}t}, {Beloborodov}, \& {den
  Hartog}}]{hascoet14ApJ}
{Hasco{\"e}t}, R., {Beloborodov}, A.~M., \& {den Hartog}, P.~R. 2014, \apjl,
  786, L1

\bibitem[{{Israel} {et~al.}(2007){Israel}, {Campana}, {Dall'Osso}, {Muno},
  {Cummings}, {Perna}, \& {Stella}}]{israel07ApJ:1647}
{Israel}, G.~L., {Campana}, S., {Dall'Osso}, S., {et~al.} 2007, \apj, 664, 448

\bibitem[{{Israel} {et~al.}(2016{\natexlab{a}}){Israel}, {Esposito}, {Rea},
  {Coti Zelati}, {Tiengo}, {Campana}, {Mereghetti}, {Rodriguez Castillo},
  {G{\"o}tz}, {Burgay}, {Possenti}, {Zane}, {Turolla}, {Perna}, {Cannizzaro},
  \& {Pons}}]{IER:2016}
{Israel}, G.~L., {Esposito}, P., {Rea}, N., {et~al.} 2016{\natexlab{a}},
  \mnras, 457, 3448

\bibitem[{{Israel} {et~al.}(2016{\natexlab{b}}){Israel}, {Esposito}, {Rea},
  {Coti Zelati}, {Tiengo}, {Campana}, {Rodriguez Castillo}, {Gotz}, {Burgay},
  {Possenti}, {Zane}, \& {Turolla}}]{israel16mnras:1935}
{Israel}, G.~L., {Esposito}, P., {Rea}, N., {et~al.} 2016{\natexlab{b}}, ArXiv
  e-prints

\bibitem[{{Israel} {et~al.}(2014){Israel}, {Rea}, {Zelati}, {Esposito},
  {Burgay}, {Mereghetti}, {Possenti}, \& {Tiengo}}]{IRZ:2014}
{Israel}, G.~L., {Rea}, N., {Zelati}, F.~C., {et~al.} 2014, The Astronomer's
  Telegram, 6370

\bibitem[{{Israel} {et~al.}(2008){Israel}, {Romano}, {Mangano}, {Dall'Osso},
  {Chincarini}, {Stella}, {Campana}, {Belloni}, {Tagliaferri}, {Blustin},
  {Sakamoto}, {Hurley}, {Zane}, {Moretti}, {Palmer}, {Guidorzi}, {Burrows},
  {Gehrels}, \& {Krimm}}]{israel08ApJ:1900}
{Israel}, G.~L., {Romano}, P., {Mangano}, V., {et~al.} 2008, \apj, 685, 1114

\bibitem[{{Kargaltsev} {et~al.}(2012){Kargaltsev}, {Kouveliotou}, {Pavlov},
  {G{\"o}{\u g}{\"u}{\c s}}, {Lin}, {Wachter}, {Griffith}, {Kaneko}, \&
  {Younes}}]{kargaltsev12apj:1834}
{Kargaltsev}, O., {Kouveliotou}, C., {Pavlov}, G.~G., {et~al.} 2012, \apj, 748,
  26

\bibitem[{{Kaspi} {et~al.}(2014){Kaspi}, {Archibald}, {Bhalerao}, {Dufour},
  {Gotthelf}, {An}, {Bachetti}, {Beloborodov}, {Boggs}, {Christensen}, {Craig},
  {Grefenstette}, {Hailey}, {Harrison}, {Kennea}, {Kouveliotou}, {Madsen},
  {Mori}, {Markwardt}, {Stern}, {Vogel}, \& {Zhang}}]{kaspi14ApJ:1745}
{Kaspi}, V.~M., {Archibald}, R.~F., {Bhalerao}, V., {et~al.} 2014, \apj, 786,
  84

\bibitem[{{Kaspi} \& {Boydstun}(2010)}]{kaspi10ApJ}
{Kaspi}, V.~M. \& {Boydstun}, K. 2010, \apjl, 710, L115

\bibitem[{{Kozlova} {et~al.}(2016){Kozlova}, {Israel}, {Svinkin}, {Frederiks},
  {Pal'shin}, {Tsvetkova}, {Hurley}, {Goldsten}, {Golovin}, {Mitrofanov}, \&
  {Zhang}}]{KIS:2016}
{Kozlova}, A.~V., {Israel}, G.~L., {Svinkin}, D.~S., {et~al.} 2016, \mnras,
  460, 2008

\bibitem[{{Kuiper} {et~al.}(2006){Kuiper}, {Hermsen}, {den Hartog}, \&
  {Collmar}}]{kuiper06ApJ}
{Kuiper}, L., {Hermsen}, W., {den Hartog}, P.~R., \& {Collmar}, W. 2006, \apj,
  645, 556

\bibitem[{{Lazarus} {et~al.}(2012){Lazarus}, {Kaspi}, {Champion}, {Hessels}, \&
  {Dib}}]{lazarus2012}
{Lazarus}, P., {Kaspi}, V.~M., {Champion}, D.~J., {Hessels}, J.~W.~T., \&
  {Dib}, R. 2012, \apj, 744, 97

\bibitem[{{Levin} {et~al.}(2010){Levin}, {Bailes}, {Bates}, {Bhat}, {Burgay},
  {Burke-Spolaor}, {D'Amico}, {Johnston}, {Keith}, {Kramer}, {Milia},
  {Possenti}, {Rea}, {Stappers}, \& {van Straten}}]{levin10ApJ}
{Levin}, L., {Bailes}, M., {Bates}, S., {et~al.} 2010, \apjl, 721, L33

\bibitem[{{Li} {et~al.}(2016){Li}, {Levin}, \& {Beloborodov}}]{li16ApJ:HW}
{Li}, X., {Levin}, Y., \& {Beloborodov}, A.~M. 2016, \apj, 833, 189

\bibitem[{{Liddle}(2007)}]{liddle07MNRAS:BIC}
{Liddle}, A.~R. 2007, \mnras, 377, L74

\bibitem[{{Lin} {et~al.}(2011){Lin}, {Kouveliotou}, {G{\"o}{\u g}{\"u}{\c s}},
  {van der Horst}, {Watts}, {Baring}, {Kaneko}, {Wijers}, {Woods}, {Barthelmy},
  {Burgess}, {Chaplin}, {Gehrels}, {Goldstein}, {Granot}, {Guiriec}, {Mcenery},
  {Preece}, {Tierney}, {van der Klis}, {von Kienlin}, \&
  {Zhang}}]{lin11ApJ:1E1841}
{Lin}, L., {Kouveliotou}, C., {G{\"o}{\u g}{\"u}{\c s}}, E., {et~al.} 2011,
  \apjl, 740, L16

\bibitem[{{Lorimer} \& {Kramer}(2012)}]{LK:2012}
{Lorimer}, D.~R. \& {Kramer}, M. 2012, {Handbook of Pulsar Astronomy}

\bibitem[{{Lyubarsky} {et~al.}(2002){Lyubarsky}, {Eichler}, \&
  {Thompson}}]{lyubarski02ApJ:magcool}
{Lyubarsky}, Y., {Eichler}, D., \& {Thompson}, C. 2002, \apjl, 580, L69

\bibitem[{{Marsden} \& {White}(2001)}]{marsden01ApJ}
{Marsden}, D. \& {White}, N.~E. 2001, \apjl, 551, L155

\bibitem[{{Mereghetti}(2008)}]{mereghetti08AARv:magentars}
{Mereghetti}, S. 2008, \aapr, 15, 225

\bibitem[{{Mereghetti} {et~al.}(2015){Mereghetti}, {Pons}, \&
  {Melatos}}]{mereghetti15:mag}
{Mereghetti}, S., {Pons}, J.~A., \& {Melatos}, A. 2015, \ssr

\bibitem[{{Ng} {et~al.}(2011){Ng}, {Kaspi}, {Dib}, {Olausen}, {Scholz},
  {G{\"u}ver}, {{\"O}zel}, {Gavriil}, \& {Woods}}]{ng11ApJ:1547}
{Ng}, C.-Y., {Kaspi}, V.~M., {Dib}, R., {et~al.} 2011, \apj, 729, 131

\bibitem[{{Pavlovi{\'c}} {et~al.}(2013){Pavlovi{\'c}}, {Uro{\v s}evi{\'c}},
  {Vukoti{\'c}}, {Arbutina}, \& {G{\"o}ker}}]{pavlovic13ApJS:SNR}
{Pavlovi{\'c}}, M.~Z., {Uro{\v s}evi{\'c}}, D., {Vukoti{\'c}}, B., {Arbutina},
  B., \& {G{\"o}ker}, {\"U}.~D. 2013, \apjs, 204, 4

\bibitem[{{Pons} \& {Rea}(2012)}]{pons12ApJ:mag}
{Pons}, J.~A. \& {Rea}, N. 2012, \apjl, 750, L6

\bibitem[{{Ransom}(2001)}]{SCOTT:2001}
{Ransom}, S.~M. 2001, PhD thesis, Harvard University

\bibitem[{{Ransom} {et~al.}(2003){Ransom}, {Cordes}, \&
  {Eikenberry}}]{RCE:2003}
{Ransom}, S.~M., {Cordes}, J.~M., \& {Eikenberry}, S.~S. 2003, \apj, 589, 911

\bibitem[{{Ransom} {et~al.}(2002){Ransom}, {Eikenberry}, \&
  {Middleditch}}]{REM:2002}
{Ransom}, S.~M., {Eikenberry}, S.~S., \& {Middleditch}, J. 2002, \aj, 124, 1788

\bibitem[{{Rea} {et~al.}(2016){Rea}, {Borghese}, {Esposito}, {Coti Zelati},
  {Bachetti}, {Israel}, \& {De Luca}}]{rea16:rcw103}
{Rea}, N., {Borghese}, A., {Esposito}, P., {et~al.} 2016, ArXiv e-prints

\bibitem[{{Rea} \& {Esposito}(2011)}]{rea11:outburst}
{Rea}, N. \& {Esposito}, P. 2011, in High-Energy Emission from Pulsars and
  their Systems, ed. D.~F. {Torres} \& N.~{Rea}, 247

\bibitem[{{Rea} {et~al.}(2013){Rea}, {Israel}, {Pons}, {Turolla}, {Vigan{\`o}},
  {Zane}, {Esposito}, {Perna}, {Papitto}, {Terreran}, {Tiengo}, {Salvetti},
  {Girart}, {Palau}, {Possenti}, {Burgay}, {G{\"o}{\u g}{\"u}{\c s}},
  {Caliandro}, {Kouveliotou}, {G{\"o}tz}, {Mignani}, {Ratti}, \&
  {Stella}}]{rea13ApJ:0418}
{Rea}, N., {Israel}, G.~L., {Pons}, J.~A., {et~al.} 2013, \apj, 770, 65

\bibitem[{{Rea} {et~al.}(2012){Rea}, {Pons}, {Torres}, \&
  {Turolla}}]{rea12ApJ:radiomag}
{Rea}, N., {Pons}, J.~A., {Torres}, D.~F., \& {Turolla}, R. 2012, \apjl, 748,
  L12

\bibitem[{{Scholz} {et~al.}(2012){Scholz}, {Ng}, {Livingstone}, {Kaspi},
  {Cumming}, \& {Archibald}}]{scholz12ApJ:1822}
{Scholz}, P., {Ng}, C.-Y., {Livingstone}, M.~A., {et~al.} 2012, \apj, 761, 66

\bibitem[{{Stamatikos} {et~al.}(2014){Stamatikos}, {Malesani}, {Page}, \&
  {Sakamoto}}]{stamatikos14:1935}
{Stamatikos}, M., {Malesani}, D., {Page}, K.~L., \& {Sakamoto}, T. 2014, GRB
  Coordinates Network, 16520

\bibitem[{{Story} \& {Baring}(2014)}]{story14ApJ}
{Story}, S.~A. \& {Baring}, M.~G. 2014, \apj, 790, 61

\bibitem[{{Str{\" u}der} {et~al.}(2001){Str{\" u}der}, {Briel}, {Dennerl},
  {Hartmann}, {Kendziorra}, {Meidinger}, {Pfeffermann}, {Reppin}, {Aschenbach},
  {Bornemann}, {Br{\" a}uninger}, {Burkert}, {Elender}, {Freyberg}, {Haberl},
  {Hartner}, {Heuschmann}, {Hippmann}, {Kastelic}, {Kemmer}, {Kettenring},
  {Kink}, {Krause}, {M{\" u}ller}, {Oppitz}, {Pietsch}, {Popp}, {Predehl},
  {Read}, {Stephan}, {St{\" o}tter}, {Tr{\" u}mper}, {Holl}, {Kemmer},
  {Soltau}, {St{\" o}tter}, {Weber}, {Weichert}, {von Zanthier},
  {Carathanassis}, {Lutz}, {Richter}, {Solc}, {B{\" o}ttcher}, {Kuster},
  {Staubert}, {Abbey}, {Holland}, {Turner}, {Balasini}, {Bignami}, {La
  Palombara}, {Villa}, {Buttler}, {Gianini}, {Lain{\' e}}, {Lumb}, \&
  {Dhez}}]{struder01aa}
{Str{\" u}der}, L., {Briel}, U., {Dennerl}, K., {et~al.} 2001, \aap, 365, L18

\bibitem[{{Sun} {et~al.}(2011){Sun}, {Reich}, {Reich}, {Xiao}, {Gao}, \&
  {Han}}]{sun11AA:SNR}
{Sun}, X.~H., {Reich}, P., {Reich}, W., {et~al.} 2011, \aap, 536, A83

\bibitem[{{Surnis} {et~al.}(2016){Surnis}, {Joshi}, {Maan}, {Krishnakumar},
  {Manoharan}, \& {Naidu}}]{surnis2016ApJ:1935}
{Surnis}, M.~P., {Joshi}, B.~C., {Maan}, Y., {et~al.} 2016, \apj, 826, 184

\bibitem[{{Szary} {et~al.}(2015){Szary}, {Melikidze}, \&
  {Gil}}]{szary15ApJ:magrad}
{Szary}, A., {Melikidze}, G.~I., \& {Gil}, J. 2015, \apj, 800, 76

\bibitem[{{Tendulkar} {et~al.}(2015){Tendulkar}, {Hasc{\"o}et}, {Yang},
  {Kaspi}, {Beloborodov}, {An}, {Bachetti}, {Boggs}, {Christensen}, {Craig},
  {Guillot}, {Hailey}, {Harrison}, {Stern}, \& {Zhang}}]{tendulkar15ApJ:0142}
{Tendulkar}, S.~P., {Hasc{\"o}et}, R., {Yang}, C., {et~al.} 2015, \apj, 808, 32

\bibitem[{{Thompson} \& {Duncan}(1996)}]{thompson96ApJ:magnetar}
{Thompson}, C. \& {Duncan}, R.~C. 1996, \apj, 473, 322

\bibitem[{{Thompson} {et~al.}(2002){Thompson}, {Lyutikov}, \&
  {Kulkarni}}]{thompson02ApJ:magnetars}
{Thompson}, C., {Lyutikov}, M., \& {Kulkarni}, S.~R. 2002, \apj, 574, 332

\bibitem[{{Torres}(2017)}]{torres17ApJ:1834}
{Torres}, D.~F. 2017, \apj, 835, 54

\bibitem[{{Turolla} {et~al.}(2015){Turolla}, {Zane}, \&
  {Watts}}]{turolla15:mag}
{Turolla}, R., {Zane}, S., \& {Watts}, A.~L. 2015, Reports on Progress in
  Physics, 78, 116901

\bibitem[{{van der Horst} {et~al.}(2012){van der Horst}, {Kouveliotou},
  {Gorgone}, {Kaneko}, {Baring}, {Guiriec}, {G{\"o}{\v g}{\"u}{\c s}},
  {Granot}, {Watts}, {Lin}, {Bhat}, {Bissaldi}, {Chaplin}, {Finger}, {Gehrels},
  {Gibby}, {Giles}, {Goldstein}, {Gruber}, {Harding}, {Kaper}, {von Kienlin},
  {van der Klis}, {McBreen}, {Mcenery}, {Meegan}, {Paciesas}, {Pe'er},
  {Preece}, {Ramirez-Ruiz}, {Rau}, {Wachter}, {Wilson-Hodge}, {Woods}, \&
  {Wijers}}]{vanderhorst12ApJ:1550}
{van der Horst}, A.~J., {Kouveliotou}, C., {Gorgone}, N.~M., {et~al.} 2012,
  \apj, 749, 122

\bibitem[{{Verner} {et~al.}(1996){Verner}, {Ferland}, {Korista}, \&
  {Yakovlev}}]{verner96ApJ:crossSect}
{Verner}, D.~A., {Ferland}, G.~J., {Korista}, K.~T., \& {Yakovlev}, D.~G. 1996,
  \apj, 465, 487

\bibitem[{{Vogel} {et~al.}(2014){Vogel}, {Hasco{\"e}t}, {Kaspi}, {An},
  {Archibald}, {Beloborodov}, {Boggs}, {Christensen}, {Craig}, {Gotthelf},
  {Grefenstette}, {Hailey}, {Harrison}, {Kennea}, {Madsen}, {Pivovaroff},
  {Stern}, \& {Zhang}}]{vogel14ApJ:2259}
{Vogel}, J.~K., {Hasco{\"e}t}, R., {Kaspi}, V.~M., {et~al.} 2014, \apj, 789, 75

\bibitem[{{Weltevrede} {et~al.}(2011){Weltevrede}, {Johnston}, \&
  {Espinoza}}]{weltevrede11MNRAS:1119}
{Weltevrede}, P., {Johnston}, S., \& {Espinoza}, C.~M. 2011, \mnras, 411, 1917

\bibitem[{{Wilms} {et~al.}(2000){Wilms}, {Allen}, \& {McCray}}]{wilms00ApJ}
{Wilms}, J., {Allen}, A., \& {McCray}, R. 2000, \apj, 542, 914

\bibitem[{{Woods} {et~al.}(2004){Woods}, {Kaspi}, {Thompson}, {Gavriil},
  {Marshall}, {Chakrabarty}, {Flanagan}, {Heyl}, \&
  {Hernquist}}]{woods04ApJ:1E2259}
{Woods}, P.~M., {Kaspi}, V.~M., {Thompson}, C., {et~al.} 2004, \apj, 605, 378

\bibitem[{{Woods} \& {Thompson}(2006)}]{woods06csxs:magnetars}
{Woods}, P.~M. \& {Thompson}, C. 2006, {Soft gamma repeaters and anomalous
  X-ray pulsars: magnetar candidates}, ed. {Lewin, W.~H.~G.~\& van der Klis,
  M.}, 547--586

\bibitem[{{Yang} {et~al.}(2016){Yang}, {Archibald}, {Vogel}, {An}, {Kaspi},
  {Guillot}, {Beloborodov}, \& {Pivovaroff}}]{yang16ApJ:1048}
{Yang}, C., {Archibald}, R.~F., {Vogel}, J.~K., {et~al.} 2016, \apj, 831, 80

\bibitem[{{Younes} {et~al.}(2015{\natexlab{a}}){Younes}, {Gogus},
  {Kouveliotou}, \& {van der Hors}}]{younes15:1935}
{Younes}, G., {Gogus}, E., {Kouveliotou}, C., \& {van der Hors}, A.~J.
  2015{\natexlab{a}}, The Astronomer's Telegram, 7213

\bibitem[{{Younes} {et~al.}(2016){Younes}, {Kouveliotou}, {Kargaltsev}, {Gill},
  {Granot}, {Watts}, {Gelfand}, {Baring}, {Harding}, {Pavlov}, {van der Horst},
  {Huppenkothen}, {G{\"o}{\u g}{\"u}{\c s}}, {Lin}, \&
  {Roberts}}]{younes16ApJ:1834}
{Younes}, G., {Kouveliotou}, C., {Kargaltsev}, O., {et~al.} 2016, \apj, 824,
  138

\bibitem[{{Younes} {et~al.}(2012){Younes}, {Kouveliotou}, {Kargaltsev},
  {Pavlov}, {G{\"o}{\v g}{\"u}{\c s}}, \& {Wachter}}]{younes12ApJ:1834}
{Younes}, G., {Kouveliotou}, C., {Kargaltsev}, O., {et~al.} 2012, \apj, 757, 39

\bibitem[{{Younes} {et~al.}(2015{\natexlab{b}}){Younes}, {Kouveliotou}, \&
  {Kaspi}}]{younes15ApJ:1806}
{Younes}, G., {Kouveliotou}, C., \& {Kaspi}, V.~M. 2015{\natexlab{b}}, \apj,
  809, 165

\end{thebibliography}
\end{document}

